\documentclass[%
 reprint,
 amsmath,amssymb,
 aps,
]{revtex4-2}
\usepackage{amsmath,amssymb,amsthm,amsfonts}
\usepackage{bm}
\usepackage{bbm}
\usepackage{braket}
\usepackage{physics}
\usepackage{mathtools}
\usepackage{graphicx}
\usepackage{hyperref}
\usepackage{enumitem}
\usepackage{dsfont}
\usepackage{comment}
\usepackage{tikz}
\usetikzlibrary{arrows.meta,calc,decorations.pathmorphing,patterns}

\def\dj{d\kern-0.4em\char"16\kern-0.1em}
\def \Dj {\mbox{\raise0.3ex\hbox{-}\kern-0.4em D}}

\def\bs{\boldsymbol}

\hypersetup{
    colorlinks=true,
    linkcolor=blue,
    citecolor=blue,
    urlcolor=blue
}

\begin{document}

\title{\bf
Entangled quantum clocks as operational probes of spacetime curvature
}

\author{Ivana \Dj or\dj evi\'{c}}
\author{Aleksandra Go\v{c}anin}
\email{aleksandra.gocanin@ff.bg.ac.rs}
\author{Dragoljub Go\v{c}anin}
\affiliation{%
Faculty of Physics, University of Belgrade\\ Studentski Trg 12-16, 11000 Belgrade, Serbia
}%

\begin{abstract}
Building on the framework developed by Perche [Phys. Rev. D 106, 025018 (2022)], we study two localized nonrelativistic quantum particles propagating along timelike geodesics in a curved spacetime background. Each particle is coupled to a quantum clock that operationally records the time spent in a prescribed spatial region. We compute the covariance of the resulting time observables for separable and entangled two-particle states, comparing flat and curved backgrounds. We then reformulate the protocol as a Bell-like experiment and show that the Bell parameter can acquire a curvature-induced correction. In particular, a protocol calibrated to saturate the classical bound in flat spacetime can be driven above this bound in curved spacetime for entangled states. We focus on two-dimensional curved backgrounds in which the local tidal term induces an effective harmonic potential in the Fermi-frame description. Our results show that spacetime curvature can modify operationally defined quantum correlations and suggest entangled quantum clocks as probes of spacetime curvature.

\end{abstract}

\maketitle

\section{Introduction}
\label{sec:intro}

One of the major lessons from string theory research is that quantum mechanics and gravity can work together consistently~\cite{GreenSchwarzWitten1987}. However, this mathematical fact generally comes at a significant cost (such as supersymmetry and extra dimensions), and it does not bring us much closer to answering some fundamental quantum gravity questions, such as how to describe the gravitational field generated by a source in a state of quantum superposition, or how quantum mechanics changes the notion of an event, or what is the quantum status of the gravitational equivalence principle~\cite{DeWittRickles2011ChapelHill, PageGeilker1981, OreshkovCostaBrukner2012,ZychBrukner2018}. 
The study of such problems in operational terms, using quantum information and quantum optics techniques, along with theoretical concepts such as quantum reference frames, in wider frameworks of quantum field theory (QFT) in curved spacetime and linearized quantum gravity, is the hallmark of relativistic quantum information (RQI). Focusing on themes such as relativistic quantum entanglement and spacetime tomography, RQI has become one of the most active research areas in fundamental physics, both theoretical and experimental~\cite{SummersWerner1985, PeresTerno2004,  Bose2017,MarlettoVedral2017,  AlsingFuentes2012,  HuLinLouko2012, ClicheKempf2010, MartinMartinezMenicucci2014, martin2011relativistic, BorregaardPikovski2025, CoveyPikovskiBorregaard2025,  GiacominiCastroRuizBrukner2019, GiacominiBrukner2022,  CastroRuiz2020, Reznik2003, SaltonMannMenicucci2015, PozasKerstjensMartinMartinez2015}.

Although QFT in curved spacetime has long offered concepts such as gravitational particle creation and Unruh/Hawking radiation, the recent development of RQI has placed new emphasis on localised, operationally defined quantum systems – wave packets, detectors and quantum clocks. In parallel, experimental research revolves around two basic challenges: implementing increasingly accurate techniques to measure the gravitational interaction between microscopic objects, and maintaining robust quantum coherence on the macroscopic scale~\cite{CarneyStampTaylor2019, BoseEtAl2025, ArndtHornberger2014, RomeroIsart2011, Westphal2021, AspelmeyerKippenbergMarquardt2014}. Together, theoretical and experimental breakthroughs are continually pushing the boundaries of our current understanding of quantum gravity.

Regarding the dynamics of non-relativistic quantum particles in a curved spacetime background, Perche established a clear criterion – the so-called \textit{Fermi bound} – for determining when such a hybrid description is appropriate~\cite{Perche2022}. Essentially, if the quantum state of a particle is sufficiently localised, one can introduce a Fermi frame around the reference timelike worldline (not necessarily geodesic) along which the particle propagates while remaining confined within a Fermi tube – a region of spacetime near the worldline that can be foliated into rest spaces supporting localised quantum states. These coordinates define an extended inertial frame, up to tidal effects encoded by components of the Riemann tensor near the worldline. The evolution of a localised wave packet in this frame is governed by an effective Hamiltonian that incorporates the reference worldline and the gravitational redshift.

In this paper, we use the Fermi frame formalism developed in ~\cite{Perche2022} to study the dynamics of a pair of entangled, non-interacting, freely falling particles in a curved spacetime background, each coupled to a Salecker–Wigner–Peres quantum clock, introduced in ~\cite{Peres1980}. The particles are confined within well-localised laboratories following their respective timelike geodesics, and the clocks measure the time each particle spends in a given spatial region inside its laboratory. These Peres-time observables are defined relationally with respect to reference geodesics, and their correlations can encode information about spacetime geometry. Focusing on two-dimensional spacetime backgrounds, especially anti-de Sitter (AdS), we first calculate the covariance of Peres-time observables for a pair of freely falling particles and discuss its dependence on the spacetime geometry and the initial state of the particles. Furthermore, we use this model to set up a Bell-like experiment and show that violation of the Clauser–Horne–Shimony–Holt (CHSH) inequality based on Peres-time observables can be interpreted as an indicator of spacetime curvature, specifically for entangled states. In this sense, entangled quantum clocks provide a means of probing spacetime geometry.
 
The paper is structured as follows. In the next section, we briefly review the one-particle formalism of ~\cite{Perche2022}. In Section III, we introduce Peres-time observables and consider a two-particle scenario, focusing on Peres-time correlations. In Section IV, we discuss the case of a two-dimensional spacetime and compare the AdS temporal correlations with those in flat spacetime. In Section V, we formulate a Bell-like protocol based on Peres-time measurements and show how curvature induces violation of the CHSH inequality for binarised Peres-time observables. Section VI presents our conclusions and outlook. Technical details are given in Appendices \ref{AppA}, \ref{AppB}, and \ref{AppC}.

\section{Dynamics of a localized particle in curved spacetime}

We start by giving a brief review of the Perche's formalism developed in~\cite{Perche2022}.
First, the formalism applies only to situations where the quantum state (wave function) of a non-relativistic mass $m$ particle is sufficiently localized with respect to the background geometry. Throughout, the particle is treated as a probe with insignificant back-reaction on spacetime. The description of such localized particle dynamics is based on the concept of a Fermi frame. The technical details of the construction can be found in Appendix \ref{AppA}. Essentially, for a given reference time-like worldline $\Gamma$ (not necessarily a geodesic) along which the particle's wave function propagates, one can introduce a Fermi coordinate system $(\tau, x^{i})$ that covers a neighborhood of $\Gamma$. Namely, the coordinate $\tau$ is the proper time along $\Gamma$, while $x^{i}$ parametrize space-like geodesics orthogonal to $\Gamma$. In these coordinates, the metric components are flat along $\Gamma$, and as we move away, they take into account the tidal corrections that are quadratic in the Euclidean distance from $\Gamma$.  

The extent to which we can shoot-out space-like geodesics from $\Gamma$ depends on the spacetime curvature and the proper acceleration of $\Gamma$. For each $\tau$, the space-like geodesics constitute a spatial slice $\Sigma_{\tau}$ that
supports a Hilbert space $\mathcal{H}_{\Sigma_\tau}\cong L^{2}(\Sigma_{\tau})$ of the propagating particle. The size of a particular $\Sigma_{\tau}$ slice is given by the Fermi bound $\ell_{\tau}$. The lowest Fermi bound $\ell_{F}$ for the foliation $\{\Sigma_{\tau}\}$ defines a Fermi tube $\mathcal{F}$ around $\Gamma$, within which the formalism is applicable.

A natural choice for the scalar product on $\Sigma_{\tau}$ is 
\begin{equation}
(\psi,\phi)_{\tau}=\int_{\Sigma_{\tau}}d\Sigma \;\psi^{*}(\boldsymbol{x})\phi(\boldsymbol{x}),  
\end{equation}
where $d\Sigma=\sqrt{g_{\Sigma}(\tau,\boldsymbol{x})}d^{n}\boldsymbol{x}$ is the invariant volume measure in $\Sigma_{\tau}$ with $g_{\Sigma}$ the determinant of the induced metric on $\Sigma_{\tau}$.
A Hamiltonian that generates the dynamics of a particle, a mapping from $\mathcal{H}_{\Sigma_\tau}$ to $\mathcal{H}_{\Sigma_{\tau+\delta\tau}}$, has to take into account the gravitational redshift because timelike worldlines inside the Fermi tube, with fixed spatial Fermi coordinates $x^{i}$, do not have the same proper time. As argued in~\cite{Perche2022}, the appropriate effective Fermi-frame Hamiltonian is given by
\begin{equation}\label{H}
\hat H
= m\mathds{1}+\frac{\hat{\boldsymbol{p}}^{2}}{2m}
+ ma_{i}(\tau)\hat x^{i}+\frac{m}{2}  R_{0i0j}(\tau) \hat x^{i}\hat x^{j}
+ \dots  ,  
\end{equation}
with coordinates and momenta operators acting on the particle's wave function as
\begin{align}
\langle\boldsymbol{x}\vert\hat{x}^{i}\vert\psi(\tau)\rangle&=x^{i} \psi(\tau,\boldsymbol{x}),\\
\langle\boldsymbol{x}\vert\hat{p}_{i}\vert\psi(\tau)\rangle&=\frac{-i}{(g_{\Sigma_{\tau}})^{\frac{1}{4}}}\frac{\partial}{\partial x^{j}}\left((g_{\Sigma_{\tau}})^{\frac{1}{4}}\psi(\boldsymbol{x})\right).
\end{align}
It can be shown that canonical commutation relations $[\hat{x}^{i},\hat{p}_{j}]=i\hbar\delta^{i}_{j}$ are indeed satisfied.

\section{Probing spacetime geometry with Peres-time correlations}
\label{sec:BP_gravity_probe}

After setting up the basic elements of the formalism for describing the dynamics of a Fermi-localized non-relativistic quantum particle in curved spacetime background, we first introduce the Peres-time observable for a single particle and then proceed to analyze the dynamics of a pair of particles and their correlations. 

\subsection{Single-particle Peres-time observable in curved spacetime background}

Consider a timelike reference worldline $\Gamma$ representing, for example, a Fermi-localized laboratory within which the particle is confined. 
The time evolution from $\tau=0$ is implemented by a unitary transformation
\begin{equation}
\hat{U}_{g,\Gamma}(\tau,0):\mathcal{H}_0 \to \mathcal{H}_\tau,
\end{equation}
generated by a (generally $\tau$-dependent) Hamiltonian
$\hat H_{g,\Gamma}$, which is a functional of the background metric and the reference worldline $\Gamma$; we will simply write $\hat{U}_{g,\Gamma}(\tau,0)\equiv \hat{U}(\tau)$. Furthermore, we assume that the particle is coupled with a quantum clock that encodes the time it spends in a spatial region $\mathcal{D}$ (e.g., a ball within the laboratory). We will consider a Salecker--Wigner--Peres (SWP) model of a quantum clock, the details of which can be found in Appendix \ref{AppB}.  

The
clock is activated only by the component of the particle's state
supported within a Fermi-fixed spatial region \(\mathcal D\). The corresponding total
Hamiltonian is
\begin{equation}
    \hat{H}_{\rm tot}
    =
    \hat{H}_{\rm p}
    +
    \hat{H}_{\rm int},
    =\hat{H}_{\rm p}+
    \lambda\hat{\Pi}_{\mathcal D}\otimes \hat{H}_{\rm c},
\end{equation}
where \(\hat{H}_{\rm p}\) is the particle Hamiltonian from (\ref{H}), \(\hat{H}_{\rm c}=
    \omega \hat{J}
    =
    - i \hbar \omega \partial_{\theta}\) is the clock
Hamiltonian, \(\lambda\) is a coupling constant and \(\hat{\Pi}_{\mathcal D}\) is the
projector into the chosen spatial region \(\mathcal D\).
The full evolution operator is 
\begin{equation}
    \hat{U}(\tau)
    =
    \mathcal T
    \exp
    \left[
    -\frac{i}{\hbar}
    \int_{0}^{\tau}
    d\tau'
    \left(
        \hat{H}_{\rm p}(\tau')
        +
        \lambda\hat{\Pi}_{\mathcal D}\otimes \hat{H}_{\rm c}
    \right)
    \right],
    \label{eq:full_particle_clock_evolution}
\end{equation}
where \(\mathcal T\) denotes time ordering.  
Since \(\hat{H}_{\rm c}\) acts only on the clock Hilbert space, we can decompose the evolution
using clock's energy eigenstates,
\begin{equation}
    \hat{H}_{\rm c}|u_{n}\rangle
    =
        n\hbar\omega |u_{n}\rangle .
\end{equation}
Then, in the \(n\)-th clock sector, the particle evolves with the effective
Hamiltonian
\begin{equation}
    \hat{H}_{\rm p}^{(n)}
    =
    \hat{H}_{\rm p}(\tau)
    +
    \lambda n\hbar\omega\hat{\Pi}_{\mathcal D}.
\end{equation}
and the full evolution operator can be written as
\begin{equation}
    \hat{U}(\tau)
    =
    \sum_n
    \hat{U}_n(\tau)
    \otimes
    |u_{n}\rangle\langle u_{n}|,
\end{equation}
where
\begin{equation}
    \hat{U}_n(\tau)
    =
    \mathcal T
    \exp
    \left[
    -\frac{i}{\hbar}
    \int_{0}^{\tau}
    d\tau'
    \left(
        \hat{H}_{\rm p}(\tau')
        +
        \lambda n\hbar\omega\hat{\Pi}_{\mathcal D}
    \right)
    \right].
\end{equation}
Each clock-energy sector sees a slightly different potential localized
in the region \(\mathcal D\), which represents the clock's backreaction on the particle.

A useful description of the particle-clock dynamics is obtained in the
interaction picture (I) with respect to the uncoupled $(\lambda=0)$ particle evolution. We denote by
\(\hat{U}_0(\tau)\) the free evolution operator that satisfies
\begin{equation}
i\hbar\frac{d}{d\tau}\hat{U}_0(\tau)
=
\hat{H}_{\rm p}(\tau)\hat{U}_0(\tau).
\end{equation}
Equivalently,
\begin{equation}
    \hat{U}_0(\tau)
    =
    \mathcal{T}\exp
    \left[
    -\frac{i}{\hbar}\int_{0}^{\tau}d\tau'
    \hat{H}_{\rm p}(\tau')
    \right].
\end{equation}
The interaction-picture
projector onto the clock-active region \(\mathcal D\) is then
\begin{equation}
    \hat{\Pi}_{\mathcal D}^{I}(\tau)
    =
    \hat{U}_0^\dagger(\tau)
    \hat{\Pi}_{\mathcal D}
    \hat{U}_0(\tau).
\end{equation}
The interaction-picture evolution operator is
\begin{equation}
    \hat{U}_{I}(\tau)
    =
    \mathcal T
    \exp
    \left[
    -\frac{i\lambda}{\hbar}
    \hat T(\tau)\otimes \hat{H}_{\rm c}
    \right],
    \label{eq:interaction_picture_evolution}
\end{equation}
where we define the Peres-time operator
\begin{equation}
    \hat T(\tau)
    =
    \int_{0}^{\tau}
    d\tau'\,
    \hat{\Pi}_{\mathcal D}^{I}(\tau').
    \label{eq:dwell_time_operator}
\end{equation}
This form makes explicit that the clock couples to the part of the
particle's state which has support in \(\mathcal D\) at each instant of evolution. 

\subsection{Weak-clock approximation}

We now consider the weak-clock approximation. The clock is
assumed to be weak in the sense that its backreaction on the particle dynamics may
be neglected. In the exact interaction picture description, the clock's evolution is governed by a time-ordered exponential (\ref{eq:interaction_picture_evolution}), since the projectors onto the clock-active region at different times need not commute. In the weak-clock regime, we approximate the the interaction picture evolution by an ordinary exponential
\begin{equation}
    \hat{U}_I(\tau)
    \simeq
    \exp
    \left[
    -\frac{i\lambda}{\hbar}
    \hat T(\tau)\otimes \hat{H}_{\rm c}
    \right].
\label{eq:weak_clock_quantum_evolution}
\end{equation}
This does not mean that the clock coupling is set to zero, nor that the accumulated clock phase is treated only to the first order in $\lambda$. 
Rather, the coherent pointer translation generated by the Peres-time operator is kept exponentiated, allowing the clock to become entangled with different dwell-time components of the particle state. What is neglected are the higher-order time-ordering corrections involving commutators of the interaction-picture projectors at different times. These terms encode the fact that successive clock-particle interactions are not independent: an earlier interaction can perturb the particle state and thereby modify its later probability of occupying the clock-active region. In the weak-clock regime, this dynamical disturbance is assumed to be negligible, while the coherent pointer shift is retained. Thus, the approximation describes a weak probe: the clock records the time spent in the selected region without appreciably altering particle’s motion.
 
For an initial product state,
the clock becomes correlated with the particle's dwell time in
\(\mathcal D\).
Let the initial particle-clock state be
\begin{equation}\label{initial}
    |\Psi_i\rangle
    =
    |\psi_i\rangle_{\text{particle}}\otimes |v_0\rangle_{\text{clock}} ,
\end{equation}
where \(|v_0\rangle=\frac{1}{\sqrt{N}}\sum_n\vert u_{n}\rangle\) is the initial clock pointer state localized around $\theta=0$; the parameter $N$ is the number of clock's energy eigenstates/marks. We now decompose
the initial particle state into 
eigenstates of the Peres-time
operator $\hat{T}(\tau)$ (note that eigenvalues/eigenvectors depend on the monitoring time $\tau$),
\begin{equation}
    |\psi_i\rangle
    =
    \sum_T c_T |T;\tau\rangle ,
    \qquad
    \hat {T}(\tau)|T;\tau\rangle
    =
T(\tau)|T;\tau\rangle .
    \label{eq:dwell_time_decomposition}
\end{equation}
Here, \(T(\tau)\) labels the possible dwell-times associated with the initial particle
state. If the spectrum is continuous, the sum should be replaced by an
integral.

Acting with Eq.~\eqref{eq:weak_clock_quantum_evolution} on the initial
state (\ref{initial}) gives us the interaction-picture, weak-clock approximated final state of the particle-clock system,
\begin{align}
&    |\Psi_f\rangle_I
=
    \hat{U}_I(\tau)
    |\Psi_i\rangle
    \nonumber\\
    &\simeq
    \exp\left[
    -\frac{i\lambda}{\hbar}
    \hat T(\tau)\otimes \hat{H}_{\rm c}
    \right]
    \sum_T c_T |T;\tau\rangle\otimes |v_0\rangle
    \nonumber\\
    &=
    \sum_T
    c_T
    |T;\tau\rangle
    \otimes
    \exp\left[
    -\frac{i\lambda}{\hbar}
    T(\tau) \hat{H}_{\rm c}
    \right]
    |v_0\rangle \nonumber\\
    &=\sum_T
    c_T
    |T;\tau\rangle
    \otimes
    |v_0(\theta-\lambda\omega T(\tau))\rangle .
\label{eq:final_quantum_dwell_time_clock_state}
\end{align}
Equation~\eqref{eq:final_quantum_dwell_time_clock_state} shows that the clock generally does not record only a single number. Instead, it becomes
entangled with the different dwell-time components of the particle state.
Each component \(|T;\tau\rangle\) shifts the clock pointer by $
    \Delta\theta_T
    =
    \lambda\omega T(\tau)$ .
In the ideal high-resolution limit, clock states corresponding to different
dwell times are distinguishable:
\begin{equation}
    \langle v_0(\theta-\lambda\omega T')
    |
    v_0(\theta-\lambda\omega T)
    \rangle
    =
    \delta_{T,T'} .
\end{equation}
Therefore, measuring the clock pointer is equivalent to measuring the eigenvalue of dwell-time \(T\). The probability of obtaining the value \(T\)
is
\begin{equation}
    P(T)
    =
    |c_T|^2
    =
    |\langle T|\psi_i\rangle|^2 .
\end{equation}
The average Peres-time is recovered as the statistical mean of the clock
outcomes:
\begin{align}
    \langle T(\tau)\rangle
    &=
     \langle \psi_i|
    \hat T(\tau)
    |\psi_i\rangle =\int_{0}^{\tau}d\tau'  \langle\psi_i| \hat{\Pi}_{\mathcal R}^{I}(\tau')  |\psi_i\rangle\nonumber \\
&=\int_{0}^{\tau}d\tau'\int_{\mathcal{{D}}} d^{n}\boldsymbol{x} \sqrt{g_{\Sigma}(\tau',\boldsymbol{x})}|\psi(\tau',\boldsymbol{x})|^{2}.
\end{align}

\subsection{Peres-time covariance}

We now consider two non-interacting particles, \(A\) and \(B\), with equal masses $m_{A}=m_{B}=m$, propagating in the same spacetime
background along prescribed laboratory worldlines, $\Gamma_A$ and $\Gamma_B$, with corresponding
proper times $\tau_A$ and $\tau_B$; each particle is coupled to a SWP clock with equal coupling constants $\lambda_{A}=\lambda_{B}=\lambda$.
We may assume that laboratory worldlines start from the same initial point and then separate.   
The total Hilbert space at the common initial Fermi slice $\Sigma_{0}$ is
\begin{equation}
\mathcal{H} = \mathcal{H}^{(A)}_0 \otimes \mathcal{H}^{(B)}_0.
\end{equation}
The Peres-time operators for the two particles $(k=A,B)$ are given by  
\begin{equation}
    \hat T_k(\tau_{k})
    =
    \int_{0}^{\tau_{k}}
    d\tau'_k
    \;\hat{\Pi}_{\mathcal{D}_k}^{I}(\tau'_k).
    \label{eq:TA_operator}
\end{equation}
Here \(\hat{\Pi}_{\mathcal{D}_{k}}^{I}(\tau_k)\) is the interaction-picture projector
onto the clock-active region \(\mathcal D_k\).

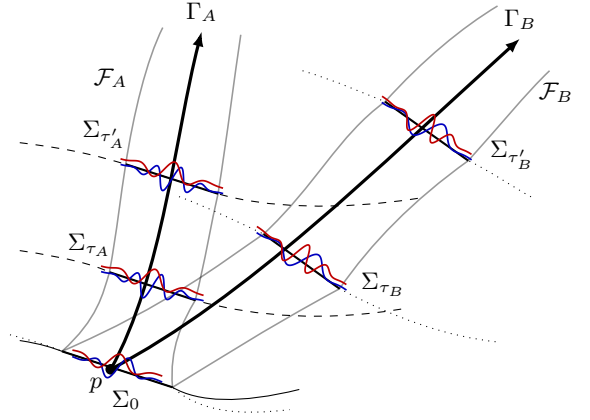
\begin{figure}[t]
\centering
\hspace*{-0.15cm}%
\begin{tikzpicture}[
    x=0.86cm,
    y=0.95cm,
    worldline/.style={very thick, -{Latex[length=2mm]}},
    tube/.style={line width=0.6pt, opacity=0.40},
    fermislice/.style={thick},
    initials/.style={thick},
    globalslice0/.style={thin},
    globalsliceA/.style={thin, dashed},
    globalsliceB/.style={thin, dotted},
    label/.style={font=\small},
    packet0blue/.pic={
        \draw[blue!70!black, line width=0.65pt, smooth, domain=-0.72:0.72, samples=110]
            plot ({\x},{0.17*exp(-7.0*\x*\x)*sin(700*\x)
                    +0.050*exp(-25*(\x-0.22)*(\x-0.22))});
    },
    packet0red/.pic={
        \draw[red!70!black, line width=0.65pt, smooth, domain=-0.72:0.72, samples=110]
            plot ({\x},{0.15*exp(-6.0*(\x+0.02)*(\x+0.02))*cos(520*\x)
                    -0.050*exp(-25*(\x+0.22)*(\x+0.22))});
    },
    packetAblue/.pic={
        \draw[blue!75!black, line width=0.65pt, smooth, domain=-0.82:0.82, samples=110]
            plot ({\x},{0.20*exp(-6.5*\x*\x)*sin(950*\x)});
    },
    packetAred/.pic={
        \draw[red!75!black, line width=0.65pt, smooth, domain=-0.82:0.82, samples=110]
            plot ({\x},{0.16*exp(-5.5*(\x+0.03)*(\x+0.03))*cos(620*\x)});
    },
    packetBblue/.pic={
        \draw[blue!75!black, line width=0.65pt, smooth, domain=-0.82:0.82, samples=110]
            plot ({\x},{0.17*exp(-5.0*\x*\x)*sin(520*\x)
                    +0.065*exp(-18*(\x-0.26)*(\x-0.26))});
    },
    packetBred/.pic={
        \draw[red!75!black, line width=0.65pt, smooth, domain=-0.82:0.82, samples=110]
            plot ({\x},{0.20*exp(-7.0*(\x-0.04)*(\x-0.04))*cos(850*\x)
                    -0.065*exp(-18*(\x+0.25)*(\x+0.25))});
    }
]

% --------------------------------------------------
% Common initial event
% --------------------------------------------------
\coordinate (O) at (1.2,0.7);
\fill (O) circle (2pt);
\node[label, below left] at (O) {$p$};

% --------------------------------------------------
% Mutual initial Fermi segment
% This segment is the common initial cross-section
% of both Fermi tubes.
% --------------------------------------------------
\coordinate (S0L) at (0.45,0.95);
\coordinate (S0R) at (2.15,0.45);

% Mutual initial global slice containing Sigma_0
% Its right wing is lowered.
\draw[globalslice0]
    (-0.2,1.10)
    .. controls (0.15,1.05) and (0.30,1.00) .. (S0L)
    -- (S0R)
    .. controls (2.55,0.24) and (3.20,0.22) .. (4.10,0.42);

% Initial B-global slice containing Sigma_0
% Its right wing is lower than the corresponding A-initial wing.
\draw[globalsliceB]
    (-0.35,1.18)
    .. controls (0.20,1.05) and (0.32,1.00) .. (S0L)
    -- (S0R)
    .. controls (2.55,0.12) and (3.15,0.05) .. (3.95,0.14);

\draw[initials] (S0L) -- (S0R);
\node[label, below] at ($(S0L)!0.58!(S0R)+(0.00,-0.13)$) {$\Sigma_{0}$};

% --------------------------------------------------
% Worldlines
% --------------------------------------------------
\draw[worldline]
    (O) .. controls (2.0,2.0) and (2.2,4.0) .. (2.6,5.4)
    node[above] {$\Gamma_A$};

\draw[worldline]
    (O) .. controls (3.3,1.7) and (5.3,3.5) .. (7.5,5.3)
    node[above] {$\Gamma_B$};

% --------------------------------------------------
% Coordinates for local Fermi slices in A tube
% These endpoints are points on opposite A-tube boundaries.
% --------------------------------------------------
\coordinate (A1L) at (1.18,2.05);
\coordinate (A1R) at (2.50,1.66);

\coordinate (A2L) at (1.43,3.55);
\coordinate (A2R) at (2.85,3.14);

% --------------------------------------------------
% Coordinates for local Fermi slices in B tube
% These endpoints are points on opposite B-tube boundaries.
% --------------------------------------------------
\coordinate (B1L) at (3.55,2.58);
\coordinate (B1R) at (4.72,1.82);

\coordinate (B2L) at (5.42,4.42);
\coordinate (B2R) at (6.70,3.60);

% --------------------------------------------------
% Fermi tube around A
% The two A-tube boundaries start at the endpoints of Sigma_0.
% --------------------------------------------------
\draw[tube]
    (S0L)
    .. controls (1.00,1.40) and (1.08,1.75) .. (A1L)
    .. controls (1.30,2.65) and (1.35,3.10) .. (A2L)
    .. controls (1.55,4.25) and (1.75,5.00) .. (2.0,5.45);

\draw[tube]
    (S0R)
    .. controls (2.10,1.00) and (2.30,1.35) .. (A1R)
    .. controls (2.65,2.20) and (2.75,2.75) .. (A2R)
    .. controls (3.00,4.00) and (3.10,4.75) .. (3.2,5.35);

\node[label] at (1.20,4.75) {$\mathcal F_A$};

% --------------------------------------------------
% Fermi tube around B
% The two B-tube boundaries also start at the endpoints of Sigma_0.
% Hence Sigma_0 fits both tubes.
% --------------------------------------------------
\draw[tube]
    (S0L)
    .. controls (2.10,1.60) and (2.85,2.15) .. (B1L)
    .. controls (4.35,3.25) and (4.70,3.95) .. (B2L)
    .. controls (6.00,5.00) and (6.60,5.45) .. (7.15,5.75);

\draw[tube]
    (S0R)
    .. controls (2.70,0.75) and (3.55,1.25) .. (B1R)
    .. controls (5.20,2.45) and (5.85,3.05) .. (B2R)
    .. controls (7.15,4.05) and (7.60,4.55) .. (7.95,4.85);

\node[label] at (8.05,4.55) {$\mathcal F_B$};

% --------------------------------------------------
% Global foliation adapted to A
% Slightly curved dashed leaves.
% Shortened on the right-hand side.
% Each A Fermi slice is a thick subsegment of the corresponding global leaf.
% --------------------------------------------------
\draw[globalsliceA]
    (-0.2,2.35)
    .. controls (0.55,2.25) and (0.95,2.15) .. (A1L)
    -- (A1R)
    .. controls (3.45,1.42) and (4.55,1.36) .. (5.75,1.55);

\draw[globalsliceA]
    (-0.2,3.85)
    .. controls (0.70,3.75) and (1.05,3.65) .. (A2L)
    -- (A2R)
    .. controls (3.55,2.95) and (4.70,2.90) .. (5.95,3.08);

% --------------------------------------------------
% Global foliation adapted to B
% Slightly curved dotted leaves.
% Shortened on the left-hand side.
% Both B global slices are shortened on the right-hand side.
% Each B Fermi slice is a thick subsegment of the corresponding global leaf.
% --------------------------------------------------
\draw[globalsliceB]
    (2.25,3.10)
    .. controls (2.70,2.95) and (3.15,2.75) .. (B1L)
    -- (B1R)
    .. controls (5.60,1.40) and (6.35,1.16) .. (7.15,1.08);

\draw[globalsliceB]
    (4.15,4.85)
    .. controls (4.70,4.70) and (5.05,4.55) .. (B2L)
    -- (B2R)
    .. controls (7.05,3.45) and (7.35,3.25) .. (7.70,3.08);

% --------------------------------------------------
% Draw local Fermi slices on top of the global slices
% --------------------------------------------------
\draw[fermislice] (A1L) -- (A1R);
\node[label, above left] at ($(A1L)!0.20!(A1R)+(-0.10,0.18)$) {$\Sigma_{\tau_A}$};

\draw[fermislice] (A2L) -- (A2R);
\node[label, above left] at ($(A2L)!0.18!(A2R)+(-0.12,0.18)$) {$\Sigma_{\tau'_A}$};

\draw[fermislice] (B1L) -- (B1R);
\node[label, right] at ($(B1R)+(0.20,0.06)$) {$\Sigma_{\tau_B}$};

\draw[fermislice] (B2L) -- (B2R);
\node[label, right] at ($(B2R)+(0.20,0.08)$) {$\Sigma_{\tau'_B}$};

% --------------------------------------------------
% Two visually distinct wave packets on each local slice.
% The initial packets have a third shape, while the later A and B
% packets have distinct evolved profiles.
% --------------------------------------------------

% Initial packets on Sigma_0
\pic[rotate=-16] at ($(S0L)!0.50!(S0R)$) {packet0blue};
\pic[rotate=-16] at ($(S0L)!0.50!(S0R)+(0.00,0.08)$) {packet0red};

% Packets centered around Gamma_A on Sigma_{tau_A}
\pic[rotate=-16] at ($(A1L)!0.50!(A1R)$) {packetAblue};
\pic[rotate=-16] at ($(A1L)!0.50!(A1R)+(0.00,0.08)$) {packetAred};

% Packets centered around Gamma_A on Sigma_{tau'_A}
\pic[rotate=-16] at ($(A2L)!0.50!(A2R)$) {packetAblue};
\pic[rotate=-16] at ($(A2L)!0.50!(A2R)+(0.00,0.08)$) {packetAred};

% Packets centered around Gamma_B on Sigma_{tau_B}
\pic[rotate=-33] at ($(B1L)!0.50!(B1R)$) {packetBblue};
\pic[rotate=-33] at ($(B1L)!0.50!(B1R)+(0.00,0.08)$) {packetBred};

% Packets centered around Gamma_B on Sigma_{tau'_B}
\pic[rotate=-33] at ($(B2L)!0.50!(B2R)$) {packetBblue};
\pic[rotate=-33] at ($(B2L)!0.50!(B2R)+(0.00,0.08)$) {packetBred};

\end{tikzpicture}
\caption{
Two-particle setup. The timelike worldlines $\Gamma_A$ and $\Gamma_B$
emanate from a common point $p$. Each worldline has its own Fermi tube, $\mathcal F_A$ and $\mathcal F_B$,
with local Fermi slices like $\Sigma_{\tau_A}$, $\Sigma_{\tau'_A}$ and
$\Sigma_{\tau_B}$, $\Sigma_{\tau'_B}$. The initial Fermi slice $\Sigma_0$ is common, and it is contained by the initial global slices of the two separate global
foliations that contain Fermi slices as their parts. The two wave profiles schematically
represent the propagation of the two localized wave packets initially prepared on $\Sigma_0$ and subsequently
evolved along the two worldlines within Fermi tubes with the corresponding Hamiltonians.
}
\label{fig:two-fermi-foliations}
\end{figure}

Let $|\psi_{i}\rangle$ be 
the initial state of the two particles, and $|v_{0, k}\rangle$ the initial state of the respective clock. The initial total state is, therefore,
\begin{equation}
    |\Psi_i\rangle
    =
    |\psi_{i}\rangle
        \otimes
    |v_{0,A}\rangle
    \otimes
    |v_{0,B}\rangle .
\end{equation}
In the weak-clock approximation, the interaction-picture evolution is
\begin{equation}
   \hat{U}_I
    \simeq
    \exp
    \left[
    -\frac{i\lambda}{\hbar}
    \left(
         \hat T_A\otimes \hat{H}_{c_A}
        +
         \hat T_B\otimes \hat{H}_{c_B}
    \right)
    \right],
\label{eq:two_clock_version3_evolution}
\end{equation}
and the final state is
\begin{align}
    |\Psi_f\rangle_I
    \simeq
    &\exp
    \left[
    -\frac{i\lambda}{\hbar}
    \left(
         \hat T_A\otimes \hat{H}_{c_A}
        +
        \hat T_B\otimes \hat{H}_{c_B}
    \right)
    \right]\nonumber\\
    &\times
    |\psi_{i}\rangle
    |v_{0,A}\rangle
    |v_{0,B}\rangle .
    \label{eq:two_clock_final_general}
\end{align}
Consider a pair of orthogonal single-particle states $\{|0\rangle, |1\rangle\}$ sufficiently localized in the initial common Fermi slice $\Sigma_{0}$, in the sense that the leakage of the tails outside $\Sigma_{0}$ (i.e., the laboratory) is negligible. The orthogonality condition can then be expressed as    
\begin{equation}
\langle 0\vert 1\rangle\simeq \int_{\Sigma_{0}}d^{n}\bs{x}\sqrt{g_{\Sigma_{0}}}\psi_{0}^{*}(\bs{x})\psi_{1}(\bs{x}).    
\end{equation}
The states $|0\rangle$ and $|1\rangle$ evolve differently for the two particles, because the two Hamiltonians, $\hat{H}_{A}$ and $\hat{H}_{B}$, drive the states along different worldlines, $\Gamma_{A}$ and $\Gamma_{B}$, from one Fermi slice to another, see Fig. \ref{fig:two-fermi-foliations}. We assume that the states remain sufficiently localized within Fermi tubes $\mathcal{F}_{A}$ and $\mathcal{F}_{B}$ around the reference worldlines, and that the (approximate) orthogonality persists for both evolutions.

Restricting the Peres-time operator to the two-dimensional subspaces spanned by $\{|0\rangle, |1\rangle\}$, we write 
\begin{align}
    \hat T_k(\tau_{k})
 &   =
    \sum_{m,n=0}^{1}
    T^k_{mn}(\tau_{k})
    |m_{k}\rangle\langle n_{k}|.   \label{eq:time_operator_matrices}
\end{align}
The matrix elements are
\begin{equation}
    T^k_{mn}(\tau_{k})
    =
    \langle m_k|\hat T_k(\tau_{k})|n_k\rangle.
\end{equation}
Since \(\hat T_k\) is Hermitian, $    T^k_{10}=(T^k_{01})^*$.
Now, the important point is that, while the orthogonality 
condition on each $\Sigma_{\tau_{k}}$ holds during evolution, the off-diagonal matrix elements need not
vanish if we restrict the integration to the clock-active regions $\mathcal{D}_{k}\subset\Sigma_{\tau_{k}}$ since
\begin{equation}
 \int_{\mathcal{D}_{k}} d^{n}\bs{x}\sqrt{g_{\Sigma}}
    \;\psi^{*}_{0}(\bs{x},\tau_{k})\psi_{1}(\bs{x},\tau_{k}) \neq 0.  
\end{equation}
We can now diagonalize the Peres-time operators:
\begin{equation}
    \hat T_k(\tau_{k})|\alpha_{k}\rangle
    =
    T_k^\alpha(\tau_{k}) |\alpha_{k}\rangle.
\end{equation}
Because \(T^k_{01}\) are nonzero, the eigenstates
\(|\alpha_{k}\rangle\) and \(|\beta_{k}\rangle\) are not the same as \(|0\rangle\) and \(|1\rangle\). 
Expanding the initial two-particle state in the Peres-time eigenbasis we get
\begin{equation}
    |\psi_i\rangle
    =
    \sum_{\alpha,\beta}
    C_{\alpha\beta}
    |\alpha_{A}\rangle|\beta_{B}\rangle .
\end{equation}
Then Eq.~\eqref{eq:two_clock_final_general} becomes
\begin{align}
 &   |\Psi_f\rangle_I
    =
    \sum_{\alpha,\beta}
    C_{\alpha\beta}
    |\alpha_{A}\rangle|\beta_{B}\rangle\nonumber\\
&    \otimes
    e^{-\frac{i \lambda}{\hbar} T_A^\alpha \hat{H}_{c_A}}
    |v_{0,A}\rangle
    e^{-\frac{i\lambda}{\hbar} T_B^\beta \hat{H}_{c_B}}
    |v_{0,B}\rangle .
    \label{eq:two_clock_final_eigenbasis}
\end{align}
For the clocks we have $
    \hat{H}_{c_k}
    =
    -i\hbar\omega
    \partial_{\theta_k}$,
and so
\begin{equation}
    e^{-\frac{i\lambda}{\hbar} T_k^\alpha \hat{H}_{c_k}}
    |v_{0,k}(\theta_k)\rangle
    =
    |v_{0,k}(\theta_k-\lambda\omega T_k^\alpha)\rangle.
\end{equation}
Thus, the final state can be written as
\begin{align}
 &   |\Psi_f\rangle_I
=
    \sum_{\alpha,\beta}
    C_{\alpha\beta}
    |\alpha_{A}\rangle|\beta_{B}\rangle\nonumber\\
&    \otimes
    |v_{0,A}(\theta_A-\lambda\omega T_A^\alpha)\rangle
    |v_{0,B}(\theta_B-\lambda\omega T_B^\beta)\rangle .
\label{eq:version3_two_observer_final_state}
\end{align}
The clocks become entangled with the
eigen-components of the Peres-time operators. Now we want to see what Peres-time correlations can tell us about the background spacetime curvature. 

\noindent
The Peres-time covariance is defined by
\begin{equation}
    {\rm Cov}(T_A,T_B)
    =
    \langle
    \hat T_A\otimes \hat T_B
    \rangle
    -
    \langle
    \hat T_A\otimes \mathbbm{1}_B
    \rangle
    \langle
    \mathbbm{1}_A\otimes \hat T_B
    \rangle ,
    \label{eq:Peres_time_covariance_def}
\end{equation}
where the expectation values are taken in the initial  state of the system. For the initial Bell state
\begin{equation}\label{Bellstate}
|\Phi^{+}\rangle=\frac{1}{\sqrt{2}}(\vert 0_{A}\rangle| 0_{B}\rangle+\vert 1_{A}\rangle| 1_{B}\rangle)    
\end{equation}
the local averages are
\begin{equation}
    \langle \hat T_k\rangle
    =
    \frac{1}{2}
    \left(
        T^k_{00}
        +
        T^k_{11}
    \right)
    \qquad
    (k=A,B).
\end{equation}
The joint expectation value is
\begin{equation}
    \langle
    \hat T_A\otimes \hat T_B
    \rangle
    =
    \frac{1}{2}
    \left(
        T^A_{00}T^B_{00}
        +
        T^A_{11}T^B_{11}
    \right)
    +
    {\rm Re}
    \left[
        T^A_{01}T^B_{01}
    \right].
\end{equation}
Therefore, the Bell state covariance is
\begin{align}
    {\rm Cov}(T_A,T_B)
    =
 &   \frac{1}{4}
    \left(
        T^A_{00}-T^A_{11}
    \right)
    \left(
        T^B_{00}-T^B_{11}
    \right)  \nonumber\\
    &+
    {\rm Re}
    \left[        T^A_{01}T^B_{01}
    \right].
    \label{eq:covariance_phi_with_coherences}
\end{align}
The second term in Eq.~\eqref{eq:covariance_phi_with_coherences} is a genuinely
quantum coherence contribution. It is present only if the restricted
Peres-time operators have nonzero off-diagonal matrix elements in the $\{|0\rangle, |1\rangle\}$ basis. On the other hand, for a  mixed state, say 
$\rho\propto(|00\rangle\langle 00|+|11\rangle\langle 11|)$,
the covariance does not have this term. Therefore, the Peres-time covariance may serve as an entanglement witness based on relationally defined (and thus diffeomorphism-invariant) temporal observables. However, this is true regardless of the spacetime geometry. In particular, this holds also for flat spacetime. We would like a more restrictive criterion, one that would make entangled states exclusively sensitive to gravity.    

\section{Peres-time correlations in (1+1)-dimensional curved spacetime}
\label{sec:fermi}

Now we can study some concrete models. Of particular interest will be the free fall of a pair of maximally entangled, non-interacting particles in (1+1)-dimensional curved spacetime. For a free fall, the single particle dynamics is governed by the Hamiltonian
\begin{equation}\label{H_geod}
\hat H_{\text{p}}
=m\mathds{1}+ \frac{\hat{\boldsymbol{p}}^{2}}{2m}
+\frac{m}{2}  R_{0i0j}(\tau) \hat x^{i}\hat x^{j},
\end{equation}
associated with the reference geodesic along which the particle propagates, and the metric components in Fermi coordinates simplify to
\begin{align}
g_{\tau\tau}&=-\left(1+R_{0i0j}(\tau) x^{i}x^{j}\right) + O(r^4),\nonumber\\
g_{\tau i}&=-\frac{2}{3}R_{0jik}(\tau)x^{j}x^{k}+\mathcal{O}(r^{4}),\nonumber\\
g_{ij}&=\delta_{ij}-\frac{1}{3} R_{ikjl}(\tau) x^{k}x^{l} + O(r^4).
\end{align} 

In two spacetime dimensions, the Riemann tensor is completely
determined by the Ricci scalar,
\begin{equation}
R_{\mu\nu\rho\sigma}
=
\frac{R}{2}
\left(
g_{\mu\rho}g_{\nu\sigma}
-
g_{\mu\sigma}g_{\nu\rho}
\right).
\end{equation}
Let $(\tau,x)$ be the Fermi coordinates constructed around a 
timelike geodesic.  
Along a geodesic ($x = 0$), the metric reduces to Minkowski metric,
\begin{equation}
g_{\mu\nu}(\tau,0) = \eta_{\mu\nu} = \mathrm{diag}(-,+),
\end{equation}
and the Riemann curvature is simply
\begin{equation}
R_{\mu\nu\rho\sigma}(\tau)
= \frac{R(\tau)}{2}
\bigl(\eta_{\mu\rho}\eta_{\nu\sigma}
    - \eta_{\mu\sigma}\eta_{\nu\rho}
\bigr),
\end{equation}
the only independent nonzero components being 
\begin{align}
R_{0x0x}&=-\frac{R(\tau)}{2}, 
\end{align}
where $R(\tau)$ is the Ricci scalar along the particle's geodesic. In general, the curvature scalar is $\tau$-dependent.
The particle dynamics is driven by a parametric harmonic oscillator Hamiltonian 
\begin{equation}
\hat{H}_{\text{p}}(\tau)
=m\mathbbm{1}+
\frac{\hat p^2}{2m}
+
\frac{1}{2}m\omega^2(\tau)\hat x^2,
\end{equation}
with $\tau$-dependent frequency  
\begin{equation}
\omega^2(\tau)
=-
\frac{R(\tau)}{2}.
\label{eq:effective_frequency_squared}
\end{equation}
For example, for (A)dS$_{2}$ spacetime, $R(\tau)=\pm 2/\ell^{2}$, 
where $\ell$ is the (A)dS radius.

The Fermi expansion of the metric components in a $(1+1)$-dimensional spacetime is simply
\begin{align}
g_{\tau\tau}&=-\left(1-\frac{R(\tau)}{2}x^{2}\right) + \mathcal{O}(x^3),\nonumber\\
g_{\tau x}&=0+\mathcal{O}(x^{3}),\nonumber\\
g_{xx}&=1+\mathcal{O}(x^3).
\end{align}
Finally, the integration measure is given by
\begin{equation}\sqrt{g_{\Sigma}}=1+\mathcal{O}\left(x^3\right),    \end{equation}
so we can regard Fermi slices as flat.

Consider a pair of localized modes $|0\rangle$ and $|1\rangle$ that propagate along a geodesic in the AdS$_{2}$ background. We may choose the first two eigenfunctions of the harmonic oscillator with the AdS frequency $\omega^{2}=\ell^{-2}$, that is
\begin{align}
\psi_{0}(x)
&= \Big(\tfrac{m\omega}{\pi\hbar}\Big)^{1/4}
  e^{-\tfrac{m\omega}{2\hbar}x^2},\\
\psi_{1}(x)&= \Big(\tfrac{m\omega}{\pi\hbar}\Big)^{1/4}
  \sqrt{\tfrac{2m\omega}{\hbar}}\, x\, 
  e^{-\tfrac{m\omega}{2\hbar}x^2}.
\end{align}
They satisfy $\langle \psi_{\alpha}\vert \psi_{\beta}\rangle=\delta_{\alpha \beta}$ when taken on the whole domain $\mathbb{R}$. These states are highly localized around the geodesic ($x=0$), 
\begin{equation}
\psi_{\alpha}(x) \sim 
\exp\!\left[-\frac{1}{2}\left(\frac{x}{L}\right)^2\right],
\end{equation}
as long as the characteristic oscillator scale $L$ is smaller than the Fermi bound, which is the AdS radius $\ell$, 
\begin{equation}
L=\sqrt{\frac{\hbar}{m\omega}}=\sqrt{\frac{\hbar\ell}{m}}<\ell,    
\end{equation}
which amounts to $m>\hbar/\ell$. If this is the case, the wavefunctions $\psi_{\alpha}$ are effectively supported within the Fermi bound.  
Hence integrating over the global slice does not conflict with localization since
outside the Fermi tube the integrand is exponentially suppressed, and so the tails of the wavefunctions that leak out from the laboratory box contribute negligibly. 

Now we define the Peres region $\mathcal{D}$ as the half-line
\begin{equation}
    \mathcal{D} = \{\, x > 0 \,\}.
\end{equation}
with $\hat{\Pi}_\mathcal{D} = \int_0^\infty |x\rangle\langle x|\,dx$ the associated 
projector, and take the time interval $[0, \tau]$. Consider a system of two freely-falling particles in AdS$_{2}$ initially in the Bell state 
$|\Phi^{+}\rangle$ given by (\ref{Bellstate}). Since each particle $(k=A,B)$ has the same frequency
$\omega_k=\omega=\ell^{-1}$, we obtain
\begin{align}
T^{(k)}_{00}&=T^{(k)}_{11}=\tau/2,\nonumber\\
T^{(k)}_{01}
&= \frac{2}{\sqrt{2\pi}\,\omega}\,
  e^{-i\omega \tau/2}\,
  \sin\!\Big(\frac{\omega \tau}{2}\Big),
\end{align}
and the covariance reads
\begin{equation}
\text{Cov}_{\text{AdS}_{2}}(T_A,T_B)
= \frac{2}{\pi \omega^2}\,
  \sin^2\!\Big(\tfrac{\omega \tau}{2}\Big)\,
  \cos(\omega \tau).
\end{equation}
In general, the Peres-time covariance is a functional of the background metric, the detector regions $\mathcal{D}_k$, and the chosen initial state. This suggests that it can be used not only as an entanglement-sensitive observable, but also, in principle, as a probe of the underlying spacetime geometry. More precisely, one may compare the measured covariance in a given background with the corresponding flat spacetime prediction obtained with the same freely-falling initial state. Deviations from the flat result then encode information about the background geometry through curvature-induced modifications of the quantum dynamics.

\begin{figure}[h!] 
    \centering
\includegraphics[width=\linewidth]{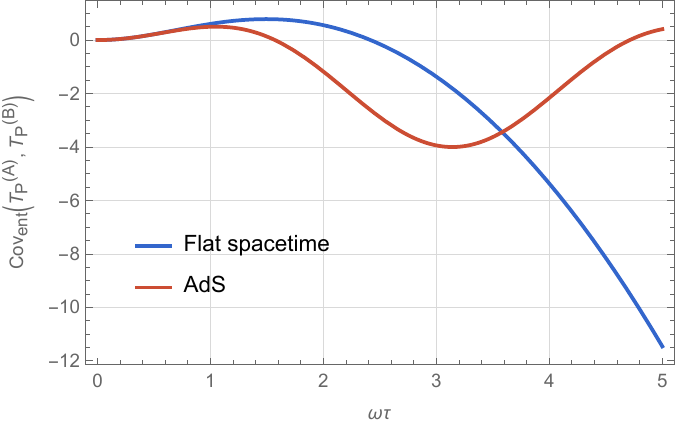}
    \caption{Peres-time covariance in flat spacetime and AdS$_2$. The two curves start to markedly differ at about $\omega\tau=1$.} 
    \label{Fig1}
\end{figure}

To illustrate this point, in Fig.~\ref{Fig1} we compare the Peres-time covariance in flat spacetime and in AdS$_2$, while keeping the same initial Bell state. In the AdS$_2$ case, these states evolve with oscillator-like phases, leading to bounded oscillatory behavior.
In flat spacetime, by contrast, the same initial profiles evolve freely under $\hat{H}_{\text{p}}=m\mathbbm{1}+\hat p^2/2m$. Then a direct calculation gives the flat spacetime covariance,
\begin{align}
\text{Cov}_{\text{flat}}(T_A,T_B)
=
&\frac{1}{2\pi\omega^2}\bigg(
\operatorname{arsinh}^2(\omega\tau)
\nonumber\\
&-
\bigg(\sqrt{1+\omega^2\tau^2}-1\bigg)^2\bigg)
.
\end{align}
In short times, $\omega\tau\ll 1$, this behaves as
\begin{equation}
\text{Cov}_{\text{flat}}(T_A,T_B)
\sim \omega^{2}\tau^{2},     
\end{equation}
and closely follows the AdS$_2$ case, but in later times the curves deviate. Thus, the qualitative difference between the flat and AdS$_2$ curves in Fig.~\ref{Fig1} arises entirely from the distinct background-dependent dynamics, since the initial states are chosen to be the same in both cases. In this sense, the Peres covariance defines a family of curvature-sensitive observables that can, at least in principle, distinguish flat from curved spacetime and probe geometric features of the background for entangled quantum states. Although a full spacetime tomography based on Peres-time correlations would require a more detailed procedure and strong experimental control, the formal structure derived here already indicates that covariance is not only an entanglement witness, but also a potential diagnostic of the geometry itself.

\section{Curvature-induced violation of a Bell-like inequality based on binarized Peres-time observables}

Now we can use the Peres-time observable to set up a Bell-like test for a pair of noninteracting freely-falling particles in curved spacetime. Since the operational meaning of the SWP clock readout has already been discussed in the previous section, we focus here on its reduction to an effective binary observable and on the resulting CHSH parameter. We consider the standard bipartite Bell scenario in which two distant observers, Alice and Bob, each have access to one member of an entangled pair and perform only local measurements in their respective laboratories. In the present setting, local measurements are realized through Peres-time observables associated with prescribed detector regions and monitoring times along the corresponding laboratory geodesics. Each observer may choose between two possible local settings, implemented by different choices of proper-time duration, and each Peres-time readout is subsequently binarized into outcomes $\pm1$. In this way, the protocol is mapped onto the usual CHSH situation, with the effective measurement settings dynamically determined by the local Hamiltonian. A detailed account of the procedure is given in Appendix \ref{AppC}.

In contrast to the free-evolution comparison used in Section IV, the Bell
construction requires a fixed reference setup in which the effective
measurement basis is defined operationally. We therefore assume that, already
in flat spacetime, each laboratory contains the same one-dimensional harmonic
trapping potential with frequency \(\omega_0\). The local settings are then
implemented by choosing different proper-time monitoring intervals.
We then place the same laboratory configuration in a curved background, still
working in the probe limit. The trapping potential is not changed by hand.
Rather, the background contributes an additional tidal term to the local
Hamiltonian. 

\subsection{Effective qubit observable from the Peres-time operator}

Let $|0\rangle$ and $|1\rangle$ be the first two harmonic-oscillator modes for frequency $\omega_{0}$,
the Peres-time operator for a given monitoring time $\tau$ in a given laboratory takes the reduced form 
\begin{equation}\hat {T}(\tau)=\begin{pmatrix}T_{00}(\tau) & T_{01}(\tau)\\T_{10}(\tau) & T_{11}(\tau)\end{pmatrix},\qquad T_{10}=T_{01}^*.
\end{equation}
Writing the off-diagonal element as
\begin{equation}
T_{01}(\tau)=|T_{01}(\tau)|e^{-i\phi(\tau)},
\end{equation}
the Peres-time operator decomposes as
\begin{equation}
\hat{T}(\tau)
=
\frac{\tau}{2}\mathbbm{1}
+
|T_{01}(\tau)|
\left[
\cos\phi(\tau)\,\sigma_x
+
\sin\phi(\tau)\,\sigma_y
\right].
\label{eq:Tdecomp}
\end{equation}
Choosing the threshold at the midpoint,
\begin{equation}
\tau^*=\frac{\tau}{2},
\end{equation}
and binarizing the Peres-time readout by classical post-processing, the corresponding binary observable is
\begin{equation}
\hat{O}(\tau)
=
\mathrm{sgn}\!\left(
\hat{T}(\tau)
-
\frac{\tau}{2}\mathbbm{1}
\right).
\end{equation}
Within the two-mode approximation this is
\begin{equation}
\hat{O}(\tau)
\simeq
\cos\phi(\tau)\,\sigma_x
+
\sin\phi(\tau)\,\sigma_y.
\label{eq:Oqubit}
\end{equation}
Thus, each proper-time duration \(\tau\) defines a measurement direction on the equator of the Bloch sphere, with the angle determined by the phase of the off-diagonal Peres matrix element.  This is the
mechanism by which the background geometry enters the Bell parameter.
In flat spacetime, we have a harmonic potential with frequency \(\omega_0\), whereas in a $(1+1)$-dimensional curved spacetime it is shifted by the curvature. Hence, the same monitoring time \(\tau\) corresponds to different effective qubit axes in the two backgrounds.

\subsection{Bell parameter}

Given two possible settings $\alpha\in\{a,a'\}$ for Alice and $\beta\in\{b,b'\}$ for Bob, the Bell parameter is
\begin{equation}
S
=
E(a,b)+E(a,b')+E(a',b)-E(a',b'),
\label{eq:CHSHdef}
\end{equation}
with correlators
\begin{equation}
E(\alpha,\beta)
=
\mathrm{Tr}\!\left[
\rho\,
\hat A_\alpha\otimes \hat B_\beta
\right]
\end{equation}
of the local observables in the equatorial form
\begin{align}
\hat A_{\alpha}
&=
\cos\phi_{\alpha}\,\sigma_x+\sin\phi_{\alpha}\,\sigma_y,\\
\hat B_{\beta}
&=
\cos\phi_{\beta}\,\sigma_x-\sin\phi_{\beta}\,\sigma_y.
\label{eq:local_observables}
\end{align}
For the Bell state
\begin{equation}
|\Phi^+\rangle
=
\frac{|00\rangle+|11\rangle}{\sqrt2},
\end{equation}
one obtains
\begin{equation}
E_{\Phi^+}(\alpha,\beta)
=
\cos(\phi_\alpha-\phi_\beta).
\end{equation}
Curvature can change the effective local axes, but it does not change the locality structure of the measurement. Hence, all separable states satisfy the CHSH bound, $|S_{\rm sep}|\leq 2$, both in flat and curved spacetime ~\cite{ClauserHorneShimonyHolt1969,Brunner2014}.

\subsection{Flat spacetime calibration}

\noindent
We now choose the flat-spacetime monitoring times so that the reference Bell operator cannot violate the CHSH bound for any input state.
The flat-spacetime Peres-time matrix is, for both particles,
\begin{align}
T_{00}&=T_{11}=\tau/2,\nonumber\\
T_{01}
&= \frac{2}{\sqrt{2\pi}\,\omega_0}\,
  e^{-i\omega_{0} \tau/2}\,
\sin\!\Big(\frac{\omega_{0} \tau}{2}\Big),
\end{align}
so the flat-spacetime  phase is 
\begin{equation}
\phi^{\rm flat}
=
\frac{\omega_0\tau}{2}.\label{flat_phase}
\end{equation}
We choose Alice's two monitoring times so that her two flat-spacetime settings coincide,
\begin{equation}
\tau_{a'}-\tau_a
=
\frac{4\pi n_A}{\omega_0}, \qquad
n_A\in\mathbb Z,
\label{eq:Alice_flat_null}
\end{equation}
that is 
\begin{equation}
\phi_{a'}^{\rm flat}-\phi_a^{\rm flat}
=
2\pi n_A \implies \hat A_a^{\rm flat}
=
\hat A_{a'}^{\rm flat}.
\end{equation}
Similarly, we choose Bob's monitoring times so that
\begin{equation}
\tau_{b'}-\tau_b
=
\frac{4\pi n_B}{\omega_0},
\qquad
n_B\in\mathbb Z,
\label{eq:Bob_flat_null}
\end{equation}
which implies
\begin{equation}
\hat B_b^{\rm flat}
=
\hat B_{b'}^{\rm flat}.
\end{equation}
With these choices, the flat-spacetime Bell operator becomes
\begin{align}
 2\hat A_a^{\rm flat}\otimes\hat B_b^{\rm flat}.
\label{eq:Bflat_reduced}
\end{align}
Since both local observables have eigenvalues \(\pm1\),
for every two-particle state \(\rho\),
\begin{equation}
|S_{\rm flat}|
\le2.
\label{eq:flat_no_violation}
\end{equation}
The flat-spacetime experiment is therefore a null Bell test: with this calibration, no state, entangled or separable, can violate the CHSH bound.

\subsection{Curvature-induced violation of Bell inequality}

We now keep the same monitoring times, but place the laboratories in a $(1+1)$-dimensional curved background.
The local harmonic trap with frequency \(\omega_0\) inside each laboratory remains unchanged,
but the tidal term modifies the effective local Hamiltonian by shifting the oscillator frequency,
\begin{equation}
\Omega^{2}_{k}(\tau)=\omega^{2}_{0}-\frac{R_{k}(\tau)}{2}.    
\end{equation}
The curved-spacetime phase entering the effective qubit observable has to be defined with
some care. The two-level system is defined by the first two oscillator modes of the reference
trap with frequency \(\omega_0\). In a curved background, however, the evolution is generated
by some $\tau$-dependent Hamiltonian. Therefore, the energy eigenstates
\(|n\rangle\) are not, in general, eigenstates of the
curved-background Hamiltonian. Even in a $(1+1)$-dimensional case, the harmonic oscillator Hamiltonian has a curvature-induced $\tau$-dependence.  
The relevant object is the off-diagonal Peres-time matrix
element
\begin{equation}
T_{nn'}^{(k)}(\tau)
=
\int_0^\tau d\tau'
\langle n|
\hat{U}_{0,k}^{\dagger}(\tau')\hat{\Pi}_{\mathcal{D}_{k}} \hat{U}_{0,k}(\tau')
|n'\rangle.
\end{equation}
Thus, in general, the curved phase is not obtained simply by replacing \(\omega_0\) with
\(\Omega_k\) in the flat-space expression (\ref{flat_phase}).

For the half-line detector region \(\mathcal{D}=\{x>0\}\), parity implies
\begin{equation}
T_{00}^{(k)}(\tau)=T_{11}^{(k)}(\tau)=\frac{\tau}{2}.
\end{equation}
Therefore, the binarized Peres-time observable remains an equatorial qubit observable. We define the curved-spacetime phase by
\begin{equation}
T_{01}^{(k)}(\tau)
=
|T_{01}^{(k)}(\tau)|e^{-i\phi_k^{\rm curv}(\tau)}.
\end{equation}
The CHSH value for the fixed state \(|\Phi^+\rangle\) is
\begin{align}
S_{\Phi^+}^{\rm curv}
=&
\cos(\phi_a^{\rm curv}-\phi_b^{\rm curv})
+
\cos(\phi_a^{\rm curv}-\phi_{b'}^{\rm curv})
\nonumber\\
\quad
+
&\cos(\phi_{a'}^{\rm curv}-\phi_b^{\rm curv})
-
\cos(\phi_{a'}^{\rm curv}-\phi_{b'}^{\rm curv}) .
\label{eq:S_curv_full}
\end{align}
This is the curved-spacetime CHSH value for the fixed monitoring times and for the
fixed initial state \(|\Phi^+\rangle\). It should be distinguished from the maximal CHSH value of the
curved Bell operator, which is obtained by optimizing over the input entangled state.
In curved spacetime, the same monitoring times generally no longer give coincident settings as in flat spacetime.
We quantify the lifting of the flat-spacetime degeneracy by
\begin{align}
\epsilon_A
&=
\left(
\phi_{a'}^{\rm curv}-\phi_a^{\rm curv}
\right)
-
\left(
\phi_{a'}^{\rm flat}-\phi_a^{\rm flat}
\right),\\
\epsilon_B
&=
\left(
\phi_{b'}^{\rm curv}-\phi_b^{\rm curv}
\right)
-
\left(
\phi_{b'}^{\rm flat}-\phi_b^{\rm flat}
\right).
\label{eq:epsilon_def}
\end{align}
These quantities measure the angular separation of settings that were identical in the flat
calibration.

\noindent
Expanding Eq.~\eqref{eq:S_curv_full}  around the flat calibration gives
\begin{equation}
S_{\Phi^+}^{\rm curv}
\simeq
2-\epsilon_A\epsilon_B,
\label{eq:S_Phi_curv_expansion}
\end{equation}
provided that the common curvature-induced shift of the two unprimed reference
axes is either absent or calibrated out by the choice of the common reference
direction, see Appendix \ref{AppC}. Therefore, the fixed input state \(|\Phi^+\rangle\) violates the
CHSH inequality whenever
\begin{equation}
\epsilon_A\epsilon_B<0.
\label{eq:Phi_violation_condition}
\end{equation}
This condition is not a restriction on the background known to Alice and Bob;
rather, it is a condition on the orientation of the two local calibrations.
It can be implemented operationally by choosing the two pairs of monitoring
times with opposite orientation.

For a constant scalar curvature \(R\), the
curvature correction to the effective angle is
\begin{align}
\delta\phi(\tau)
=
&-\frac{R}{4\omega_0^2}
\left[
\frac{\omega_0\tau}{2}
-
\frac{
\sin(\omega_0\tau)
\left[
2+\cos(\omega_0\tau)
\right]
}{6}
\right]\nonumber\\
&+
\mathcal{O}\left(
R^2/\omega_0^4
\right).
\label{eq:delta_phi_constant_curvature_CHSH}
\end{align}
Using the calibrated time differences, 
the periodic part of Eq.~\eqref{eq:delta_phi_constant_curvature_CHSH}
cancels, and one finds
\begin{equation}
\epsilon_k
=
-\frac{\pi }{2\omega_0^2}n_k R_k
+
\mathcal{O}\left(
R_k^2/\omega_0^4
\right).
\label{eq:epsilon_A_scalar_curvature}
\end{equation}
As shown in Appendix \ref{AppC}, one obtains
\begin{equation}
S_{\Phi^+}^{\rm curv}
=
2
-
\frac{\pi^2}{4\omega_0^4}
 n_A n_B R_A R_B
+
\mathcal{O}(R^3).
\label{eq:CHSH_scalar_curvature2}
\end{equation}
The fixed-state violation condition is therefore
\begin{equation}
n_A n_B R_A R_B<0.
\label{eq:violation_scalar_condition}
\end{equation}
The sign of \(n_A\) or \(n_B\) is determined by the ordering of the two
monitoring times at the corresponding laboratory. A negative value of
\(n_B\), for example, means that $
\tau_{b'}<\tau_b$,
while both monitoring times remain positive.

In (A)dS$_{2}$,
\begin{equation}
R_A
=
R_B
=
\pm\frac{1}{2\ell^{2}}.
\end{equation}
Therefore the violation condition becomes
\begin{equation}
n_A n_B<0.
\end{equation}
This can be achieved by choosing the two local flat calibrations with
opposite orientation, for example
\begin{equation}
\tau_{a'}-\tau_a
=
\frac{4\pi |n_A|}{\omega_0},
\qquad
\tau_{b'}-\tau_b
=
-\frac{4\pi |n_B|}{\omega_0},
\end{equation}
with all monitoring times kept positive. Then
\begin{equation}
S_{\Phi^+}^{\rm AdS}
\simeq
2
+
\frac{\pi^2 }{\ell^4\omega_0^4}|n_A n_B|.
\label{eq:S_Phi_AdS}
\end{equation}
The result is operational. The initial state, the four monitoring times, the
clock-active regions, and the binarization rule are fixed before the
experiment. Alice and Bob do not adjust the protocol after learning the
background geometry. They repeatedly prepare the same state \(|\Phi^+\rangle\),
select between the preassigned local monitoring times, record the binary clock
outcomes, and estimate the four correlators entering
Eq.~\eqref{eq:S_curv_full}. In flat spacetime the same protocol
gives $S_{\Phi^+}^{\rm flat}=2$,
whereas in a constant-curvature background it gives the curvature-dependent
value in Eq.~\eqref{eq:CHSH_scalar_curvature2}. Therefore, the same initial Bell
state, the same local clocks, and the same pre-agreed monitoring times produce
no CHSH violation in the flat calibrated setup, but can produce a violation in
a curved background. The violation is inferred directly from the measurement
statistics, without Alice and Bob needing to know the background geometry in
advance. The curvature can be obtained operationally from the
measured value of the Bell parameter.

The absence of a linear curvature correction is a direct consequence of the
flat-space degeneracy imposed on both local pairs of measurements. Curvature
must lift the degeneracy on Alice's and Bob's sides simultaneously before the
fixed-state CHSH value can move away from the calibrated flat value. The
resulting Bell activation is therefore bilinear in the two local curvature
responses and quadratic in a common weak-curvature scale. This reduces
metrological sensitivity, but provides a clean null-test interpretation: the
calibrated flat-space protocol gives \(S^{\text{flat}}_{\Phi^+}=2\), whereas a curved
background can activate a violation  \(S_{\Phi^+}^{\rm curv}>2\). In this
operational sense, the measured Bell parameter acts as a witness of the
background curvature. 
Moreover, mixed states cannot change the flat-space value even in a curved spacetime, and so it is the cooperation between curvature and entanglement that enables violation.  
The protocol therefore provides a curvature-induced Bell witness based the dynamical map between proper-time duration and effective measurement
axis, encoded in the phase of the off-diagonal Peres-time matrix element.

\section{Conclusion and outlook}

In this work, we apply the Fermi frame formalism from \cite{Perche2022} to study a pair of non-interacting quantum particles propagating along timelike geodesics in a curved background. Each particle carries an auxiliary system – a quantum clock – that records the corresponding dwell time in a specified spatial region. From these relationally defined local Peres-time observables, we construct joint temporal correlations and discuss how they depend on the initial state and on the background geometry. Our first result is that Peres-time covariance distinguishes coherent entanglement from classical correlations using only local measurements. The covariance thus provides an operational entanglement witness that can also serve as a measure of the deviation from flat geometry.

Our main result is based on a Bell-like formulation of two particle-clock systems. After a binarization of the Peres-time observable at the midpoint of the monitoring interval, the restricted Peres-time operators define effective equatorial qubit measurements. Their measurement axes are dynamically determined by the phases of the off-diagonal Peres-time matrix elements. The resulting experiment has the structure of a CHSH test, with local settings implemented through monitoring times, clock readouts, and spatial filters rather than through conventional spin analyzers.
The monitoring times are calibrated in flat spacetime so that the two local settings on each side coincide. The corresponding CHSH operator is then degenerate and cannot violate the classical bound for any input state (entangled or separable). The same monitoring times are subsequently used without modification in a curved background. Curvature changes the local dynamics and, consequently, the map between the monitoring time and effective measurement axis. This lifts the flat-space degeneracy. In the weak-curvature regime, the local angular splittings are linear in the tidal curvature, while the leading Bell activation is quadratic in the tidal curvature. For the fixed input Bell state and the appropriate preassigned orientation of the monitoring-time pairs, the same protocol that gives $S=2$ in flat spacetime can give $S>2$ in curved spacetime. The Bell parameter therefore functions as an operational witness of curvature: the local measurement procedure, the initial state, and the classical CHSH bound remain unchanged, while the observed statistics reveals the curvature-dependence.

The construction is local and perturbative. Its validity requires weak curvature relative to the trapping scale and localization within the Fermi tube. In this regime, the quadratic curvature dependence of the CHSH correction agrees with the tidal Hamiltonian. Local measurement-axis shifts are linear in curvature, and the violation requires that the degeneracy be lifted on both sides.

\medskip

Several extensions seem natural. The state dependence should be studied beyond maximally entangled pure states. It would also be interesting to consider different spacetime backgrounds with time-dependent or spatially varying curvature. In particular, we could adapt the formalism to a pair of free-falling particles in a Schwarzschild background. Furthermore, keeping the AdS case in mind, one could perhaps make a connection to the boundary description in the context of AdS/CFT duality. Non-geodesic laboratories and interaction between the particles provide further generalizations.

\medskip

Another question is how much geometric information can be reconstructed from families of temporal observables. Measurements performed with different monitoring intervals, detector regions, and initial states may probe different curvature components. This suggests a possible route towards operational spacetime tomography based on local quantum clocks and correlation measurements. Establishing the attainable precision, identifying the dominant systematic effects, and connecting the protocol with realistic atomic, trapped-particle, optomechanical, or detector-based platforms remain important directions for future work.

\section{Acknowledgement}
\label{sec:acknowledgement}
The authors thank \v{C}aslav Brukner for helpful discussions and suggestions. This work is supported by funding provided by the Faculty of Physics University of Belgrade, through grant number 451-03-47/2023-01/200162 from the Ministry of Science, Technological Development and Innovations of the Republic of Serbia. The authors acknowledge the support of the Science Fund of the Republic of Serbia, grant number 9029-YF-SAIGE, Twisted Holography: A Holographic Stance on the Quantum Superposition of Spacetimes (HOLISTIQUS).

\bibliographystyle{apsrev4-2}
\bibliography{ref.bib}

\appendix

\section{Fermi-frame construction}
\label{AppA}

Given a reference time-like worldline $\Gamma$ (not necessarily a geodesic) with parametric equations $z^{\mu}=z^{\mu}(\tau)$, where $\tau$ is the proper time parameterizing $\Gamma$, we can set up Fermi normal coordinates $(\tau, x^{i})$ in the vicinity of $\Gamma$ as follows. 
We first introduce a local orthonormal frame $\boldsymbol{e}_{\mu}(\tau_{0})$ in the tangent space to a given
point $z^{\mu}(\tau_{0})$ of the worldline, satisfying $\boldsymbol{e}^{\mu}_{0}(\tau_{0})=u^{\mu}$, where $u^{\mu}$ is the $4$-velocity of the worldline $\Gamma$. The orthonormality condition means that $g(\boldsymbol{e}_{\mu},\boldsymbol{e}_{\nu})=\eta_{\mu\nu}=\text{diag}(-1,1,1,1)$. Starting from the initial point $z^{\mu}(\tau_{0})$ we use Fermi transport to extend the orthonormal frame $\boldsymbol{e}_{a}(\tau_{0})$ along $\Gamma$,
\begin{equation}
\frac{D (e_{\mu})^{\alpha}}{d\tau}+2a^{[\alpha}u^{\beta]}(e_{\mu})_{\beta}=0,    
\end{equation}
where $D/d\tau$ is the directional
covariant derivative along $z^{\mu}(\tau)$ and $a^{\mu}=du^{\mu}/d\tau$ is the $4$-acceleration of $\Gamma$. This operation is a generalization of a more common parallel transport along a curve. In particular, $4$-velocity is parallel-transported along a geodesic, but it is always Fermi-transported along a timelike worldline, because Fermi transport takes into account the curvature of the worldline when transporting vectors from one tangent space to another. The Fermi-transported family of orthonormal frames $\boldsymbol{e}_{\mu}(\tau)$ along $z^{\mu}(\tau)$, where $e^{\mu}_{0}(\tau)=u^{\mu}(\tau)$ for all $\tau$, constitutes the Fermi frame. 

Now, to introduce spatial Fermi coordinates $x^{i}$ we pick a point $\boldsymbol{z}(\tau)$ from the worldline and follow spacelike geodesics with tangent vectors orthogonal to $u^{\mu}(\tau)$. The set of all points that can be reached in this way by a unique geodesic constitute a local rest space $\Sigma_{\tau}$. The family of local rest spaces realizes a foliation $\mathcal{F}=\cup_{\tau}\Sigma_{\tau}$, of the spacetime region around $\Gamma$. Spatial coordinates are assigned to a point $p\in\Sigma_{\tau}$ using the exponential map $x^{i}_{p}=\text{exp}_{\boldsymbol{z}(\tau)}(x^{i}e_{i}(\tau))$. Thus, Fermi normal coordinates parametrize the whole region $\mathcal{F}$. The proper distance from the worldine is given by the Euclidean norm of the spatial Fermi coordinates, $r=\sqrt{\delta_{ij}x^{i}x^{j}}$. Finally, we can extend the Fermi frame to the entire $\mathcal{F}$ by parallel-transporting along the spacelike geodesics that make up $\Sigma_{\tau}$ slices. The metric components in Fermi normal coordinates have a particularly nice form given by
\begin{align}\label{Fermi_metric}
g_{\tau\tau}&=-(1+a_{i}(\tau)x^{i})^{2}-R_{0i0j}(\tau) x^{i}x^{j} + O(r^3),\nonumber\\
g_{\tau i}&=-\frac{2}{3}R_{0jik}(\tau)x^{j}x^{k}+\mathcal{O}(r^{3}),\nonumber\\
g_{ij}&=\delta_{ij}-\frac{1}{3} R_{ikjl}(\tau) x^{k}x^{l} + O(r^3),
\end{align}
where $a^{\mu}(\tau)$ and $R_{\mu\nu\alpha\beta}(\tau)$ denote the components of acceleration and curvature in the Fermi coordinates at $z^{\mu}(\tau)$.
This expansion is valid if $r$ is sufficiently smaller than
both the curvature radius of spacetime and $1/a$, where
$a=\sqrt{a_{\mu}a^{\mu}}$ is the magnitude of the proper acceleration of
the worldline.

For each slice $\Sigma_{\tau}$ one defines the $\tau$-Fermi bound $\ell_{\tau}$ defined as the minimum proper length of the maximally extended geodesics in $\Sigma_{\tau}$. The Fermi bound is then defined as $\ell_{F}=\text{inf}_ {\tau} \;\ell_{\tau}$. Each of the $\tau$-Fermi bounds defines a bound for the size of
a system in $\Sigma_{\tau}$ which can be described in terms of Fermi normal coordinates. Thus, the Fermi-bound is
a bound for the size of a system centered at the reference worldline $\Gamma$
that can be entirely described by Fermi normal coordinates at all times. The Fermi bound also defines a world
tube around $\Gamma$, where systems that can be
fully described in the Fermi frame may have
support. This tube is defined as the region - Fermi tube - spanned by
all geodesics contained in $\Sigma_{\tau}$ that have a proper length
smaller than the Fermi bound $\ell_{F}$ for each $\tau$.
This estimate can be
useful for providing bounds for the regime of validity of
frameworks which use Fermi normal coordinates. It is argued in ~\cite{Perche2022} that 
the Fermi bound can be estimated from below as 
\begin{equation}
\ell_{F}\gtrsim \text{inf}_{\tau}\left(\frac{1}{a+\sqrt{\lambda_{R}}}\right),    
\end{equation}
where $a=\sqrt{a_{\mu}a^{\mu}}$ is the magnitude of the $4$-acceleration of $z^{\mu}(\tau)$, and $\lambda_{R}$ is the largest eigenvalue of the matrix $-R_{0i0j}$, if there are any. For a geodesic, $a=0$.

\section{Salecker-Wigner-Peres clock}\label{AppB}

Here we briefly recall the SWP clock, a simple model of a
quantum clock useful for defining elapsed time in quantum mechanics.
The clock is introduced because in quantum mechanics time is not represented by
an operator. Instead, we infer the elapsed time from the state of an auxiliary quantum system coupled to
a particle.
We assume that the clock has an odd number of states, $N = 2j+1$.
The clock Hamiltonian is
\begin{equation}
\hat{H}_{c}
    =
    \omega \hat{J}
    =
    - i \hbar \omega \frac{\partial}{\partial \theta},
\end{equation}
where \(\theta\) describes the position of the clock's hand. The eigenstates
of the clock are
\begin{equation}
    u_n(\theta)=\langle\theta|u_{n}\rangle
    =
    \frac{1}{\sqrt{2\pi}} e^{i n \theta},
\end{equation}
and satisfy
\begin{equation}
    \hat{H}_{c} \vert u_n\rangle
    =
    n \hbar \omega| u_n\rangle.
\end{equation}
Their time evolution is
\begin{equation}
    e^{- i \hat{H}_{c} \tau/\hbar} |u_n\rangle
    =
    e^{-in\omega \tau}|u_n\rangle,
\end{equation}
where $\tau$ is the Fermi time coordinate. 

It is useful to introduce the localized clock basis (pointer states)
\begin{equation}
    |v_s\rangle
    =
    \frac{1}{\sqrt{N}}
    \sum_{n=-j}^{j}
    e^{-2\pi i s n/N} |u_n\rangle,
\end{equation}
with $s=0,\ldots,N-1$,
or, equivalently,
\begin{equation}
    v_s(\theta)
    =
    \frac{1}{\sqrt{2\pi N}}
    \frac{
    \sin\!\left[ N(\theta-2\pi s/N)/2 \right]
    }{
    \sin\!\left[ (\theta-2\pi s/N)/2 \right]
    } .
\end{equation}
For large \(N\), the state $|v_s\rangle$ is peaked around
\begin{equation}
    \theta = \frac{2\pi s}{N},
\end{equation}
with angular uncertainty of order \(\pm \pi/N\). Thus, $|v_s\rangle$ represents the state in which
clock's hand points to the \(s\)-th hour. 
%Now we have\begin{equation} e^{- i H_{\rm clock} \tau/\hbar}| v_s\rangle =e^{-in\omega t}|v_s\rangle.\end{equation}
If \(\tau\) is such that
\begin{equation}
   \omega \tau = \frac{2\pi}{N},
\end{equation}
then
\begin{equation}
    e^{- i \hat{H}_{c} \tau/\hbar} |v_s\rangle
    =
    |v_{s+1 \,({\rm mod}\, N)}\rangle,
\end{equation}
which justifies the clock interpretation.

\section{Curvature-induced violation of the Peres-time Bell inequality}
\label{AppC}

Here we give the explicit calculation of the Peres-time matrix in a weakly-curved \((1+1)\)-dimensional background and show how a curvature-induced Bell violation arises when the flat-spacetime calibration is reused in a curved spacetime.

We consider a localized particle confined in a laboratory with a harmonic trapping potential of
frequency \(\omega_0\). In two spacetime dimensions the Riemann tensor is
completely determined by the Ricci scalar,
\begin{equation}
R_{\mu\nu\rho\sigma}
=
\frac{R}{2}
\left(
g_{\mu\rho}g_{\nu\sigma}
-
g_{\mu\sigma}g_{\nu\rho}
\right).
\end{equation}
In the Fermi frame associated with the laboratory geodesic, using signature
\((-+)\), one has
\begin{equation}
R_{0x0x}(\tau)
=
-\frac{1}{2}R(\tau).
\end{equation}
The leading tidal correction to the nonrelativistic particle Hamiltonian is therefore
quadratic in the spatial Fermi coordinate,
\begin{equation}
\hat H_{\text{p}}(\tau)
=m\mathbbm{1}
+\frac{\hat p^2}{2m}
+
\frac{1}{2}m\omega_0^2\hat x^2
-
\frac{m}{4}R(\tau)\hat x^2 .
\label{eq:curved_time_dependent_H}
\end{equation}
Equivalently,
\begin{equation}
\hat H_{\text{p}}(\tau)
=m\mathbbm{1}
+
\frac{\hat p^2}{2m}
+
\frac{1}{2}m\Omega^2(\tau)\hat x^2,
\end{equation}
where
\begin{equation}
\Omega^2(\tau)
=
\omega_0^2
-
\frac{1}{2}R(\tau)
\label{eq:effective_frequency_squared}
\end{equation}
is the curvature-modified frequency. 

\subsection{Weak-curvature dynamics in $(1+1)$-dimensions}

\noindent
In the weak-curvature regime,
\begin{equation}
\frac{|R(\tau)|}{\omega_0^2}\ll 1,
\label{eq:weak_curvature_regime}
\end{equation}
the effective frequency is
\begin{equation}
\Omega(\tau)
=
\sqrt{\omega_0^2-\frac{1}{2}R(\tau)}
\simeq
\omega_0
-
\frac{R(\tau)}{4\omega_0}.
\label{eq:effective_frequency_expansion}
\end{equation}
It is useful to split the particle Hamiltonian into the reference oscillator Hamiltonian
and a curvature-dependent perturbation,
\begin{equation}
\hat H_{\text{p}}(\tau)=\hat H_0+\hat V(\tau),
\end{equation}
where
\begin{equation}
\hat H_0
=m\mathbbm{1}
+
\frac{\hat p^2}{2m}
+
\frac{1}{2}m\omega_0^2\hat x^2,
\label{eq:reference_Hamiltonian}
\end{equation}
and
\begin{equation}
\hat V(\tau)
=
-\frac{m}{4}R(\tau)\hat x^2.
\label{eq:curvature_perturbation}
\end{equation}
The eigenstates of \(\hat H_0\) satisfy (we can ignore the constant $m$ shift)
\begin{equation}
\hat H_0|n\rangle
=
E_n^{(0)}|n\rangle,
\qquad
E_n^{(0)}
=
\hbar\omega_0
\left(
n+\frac{1}{2}
\right).
\end{equation}
Using
\begin{equation}
\hat x
=
\sqrt{\frac{\hbar}{2m\omega_0}}
\left(
\hat a+\hat a^\dagger
\right),
\end{equation}
the perturbation becomes
\begin{equation}
\hat V(\tau)
=
-\frac{\hbar R(\tau)}{8\omega_0}
\left(
\hat a^2
+
\hat a^{\dagger 2}
+
2\hat N
+
1
\right),
\label{eq:curvature_perturbation_ladder}
\end{equation}
where
\begin{equation}
\hat N=\hat a^\dagger\hat a .
\end{equation}
The diagonal term \(2\hat N+1\) shifts the phases of the oscillator levels.
The off-diagonal terms \(\hat a^2\) and \(\hat a^{\dagger 2}\) mix levels with
\(\Delta n=\pm2\).

In the interaction picture with respect to \(\hat H_0\),
\begin{equation}
\hat a_I(s)
=
\hat a e^{-i\omega_0 s},
\qquad
\hat a_I^\dagger(s)
=
\hat a^\dagger e^{i\omega_0 s}.
\end{equation}
Hence
\begin{equation}
\hat V_I(s)
=
-\frac{\hbar R(s)}{8\omega_0}
\left(
\hat a^2 e^{-2i\omega_0s}
+
\hat a^{\dagger 2}e^{2i\omega_0s}
+
2\hat N
+
1
\right).
\label{eq:interaction_picture_perturbation}
\end{equation}
To first order in the curvature,
\begin{equation}
\hat U_I(\tau)
=
\mathbbm{1}
-
\frac{i}{\hbar}
\int_0^\tau ds\,\hat V_I(s)
+
\mathcal O(R^2).
\label{eq:first_order_interaction_evolution}
\end{equation}
The diagonal phase correction is controlled by
\begin{equation}
\mathcal R_0(\tau)
=
\int_0^\tau ds\,R(s),
\label{eq:integrated_curvature}
\end{equation}
whereas leakage outside the original oscillator subspace is controlled by
\begin{equation}
\mathcal R_2(\tau)
=
\int_0^\tau ds\,R(s)e^{2i\omega_0s}.
\label{eq:resonant_curvature_component}
\end{equation}
For the initial ground state, the Schrödinger-picture evolution is
\begin{align}
|\psi_0(\tau)\rangle
={}&
e^{-i\omega_0\tau/2}
\left[
1+
\frac{i}{8\omega_0}
\mathcal R_0(\tau)
\right]|0\rangle
\nonumber\\
&+
\frac{i\sqrt{2}}{8\omega_0}
e^{-i5\omega_0\tau/2}
\mathcal R_2(\tau)
|2\rangle
+
\mathcal{O}(R^2).
\label{eq:ground_state_curved_evolution}
\end{align}
Similarly, for the first excited state,
\begin{align}
|\psi_1(\tau)\rangle
={}&
e^{-i3\omega_0\tau/2}
\left[
1+
\frac{3i}{8\omega_0}
\mathcal R_0(\tau)
\right]|1\rangle
\nonumber\\
&+
\frac{i\sqrt{6}}{8\omega_0}
e^{-i7\omega_0\tau/2}
\mathcal R_2(\tau)
|3\rangle
+
\mathcal{O}(R^2).
\label{eq:first_excited_curved_evolution}
\end{align}
Equations~\eqref{eq:ground_state_curved_evolution} and
\eqref{eq:first_excited_curved_evolution} show that curvature produces both
phase shifts and leakage outside the original energy state. 
The phase-only approximation is valid provided that the off-diagonal
curvature-induced mixing is negligible. A sufficient condition is
\begin{equation}
\frac{|\mathcal R_2(\tau)|}{\omega_0}\ll1.
\label{eq:small_leakage_condition}
\end{equation}
This condition excludes resonant curvature profiles with appreciable Fourier
weight near \(2\omega_0\). For a slowly varying curvature profile, the
oscillatory factor \(e^{2i\omega_0s}\) averages the leakage amplitude to a
small value, and the dominant effect is the diagonal phase shift. If \(R(\tau)\) contains a component
near \(2\omega_0\), the terms \(\hat a^2\) and \(\hat a^{\dagger 2}\) can drive transitions $|n\rangle\longrightarrow |n\pm2\rangle$,
and the two-level phase-only description is no longer sufficient.

In the remainder, we focus on constant curvature, where the
Hamiltonian is time independent, and the Bell activation can be written
explicitly.

\subsection{Constant curvature}

\noindent
Now we consider the constant curvature case,
\begin{equation}
R(\tau)=R_0 .
\end{equation}
Then the Hamiltonian is time independent,
\begin{equation}
\hat H_{\text{p}}
=m\mathbbm{1}
+
\frac{\hat p^2}{2m}
+
\frac{1}{2}m\Omega^2\hat x^2,
\end{equation}
with 
\begin{equation}
\Omega^2
=
\omega_0^2
-
\frac{R_0}{2}.
\label{eq:constant_curvature_frequency}
\end{equation}
Thus, constant curvature simply shifts the oscillator frequency. In the weak-curvature regime,
\begin{equation}
\Omega
=
\omega_0
-
\frac{R_0}{4\omega_0}
+
\mathcal O\left(\frac{R_0^2}{\omega_0^3}\right).
\label{eq:constant_curvature_frequency_expansion}
\end{equation}
If the Hamiltonian is expressed in the original \(\omega_0\)-oscillator basis,
the perturbation contains both diagonal and off-diagonal terms. 
For constant curvature,
\begin{equation}
\mathcal R_0(\tau)
=
R_0\tau,
\end{equation}
and
\begin{equation}
\mathcal R_2(\tau)
=
R_0
\frac{
e^{2i\omega_0\tau}-1
}{
2i\omega_0
}.
\end{equation}
The clock-active region is chosen to be the half-line
\begin{equation}
\mathcal D=[0,\infty),
\end{equation}
with projector
\begin{equation}
\hat\Pi_D
=
\int_0^\infty dx\,|x\rangle\langle x|.
\end{equation}
The Peres-time operator associated with this region is
\begin{equation}
\hat T(\tau)
=
\int_0^\tau ds\,\hat\Pi_D^{I}(s),
\label{eq:Peres_operator_integrated}
\end{equation}
and its matrix elements are
\begin{equation}
T_{mn}(\tau)
=
\int_0^\tau ds\,\Pi_{mn}^{I}(s).
\label{eq:Tmn_integrated}
\end{equation}
For the half-line region, parity implies
\begin{equation}
\Pi_{00}^{I}(s)
=
\Pi_{11}^{I}(s)
=
\frac{1}{2}
\end{equation}
at every instant. Therefore,
\begin{equation}
T_{00}(\tau)
=
T_{11}(\tau)
=
\frac{\tau}{2}.
\label{eq:T_diagonal_integrated}
\end{equation}

The off-diagonal element determines the effective measurement axis. For weak
constant curvature one obtains
\begin{align}
&T_{01}(\tau)
={}
\frac{1}{\sqrt{2\pi}}
\int_0^\tau ds\,e^{-i\omega_0s}
\\
&+
\frac{iR_0}{4\omega_0\sqrt{2\pi}}
\int_0^\tau ds\,e^{-i\omega_0s}
\left[
s
-
\frac{\sin(2\omega_0s)}{2\omega_0}
\right]
+
\mathcal O(R_0^2).\nonumber
\label{eq:T01_constant_integral}
\end{align}
Writing the off-diagonal element in polar form,
\begin{equation}
T_{01}(\tau)
=
|T_{01}(\tau)|e^{-i\phi(\tau)},
\label{eq:T01_polar}
\end{equation}
one finds, away from the isolated zeros of the flat off-diagonal element,
\begin{align}
\phi(\tau)
={}&
\frac{\omega_0\tau}{2}
-
\frac{R_0}{4\omega_0^2}
\left[
\frac{\omega_0\tau}{2}
-
\frac{
\sin(\omega_0\tau)
\left[
2+\cos(\omega_0\tau)
\right]
}{6}
\right]
\nonumber\\
&+
\mathcal O\left(
R_0^2/\omega_0^4
\right).
\label{eq:integrated_effective_angle_R}
\end{align}
The first term in the square brackets is the contribution due to the
curvature-induced frequency shift. The oscillatory term is the first-order
effect of the mismatch between the \(\omega_0\) and
\(\Omega\)-oscillator bases. 

The Peres-time matrix reduced to
$
\mathcal H_2=\mathrm{span}\{|0\rangle,|1\rangle\}$
is therefore
\begin{equation}
\hat T(\tau)=
\begin{pmatrix}
\tau/2 & T_{01}(\tau) \\
T_{01}^*(\tau) & \tau/2
\end{pmatrix}.
\label{eq:reduced_T_matrix_general}
\end{equation}
Using Eq.~\eqref{eq:T01_polar}, this becomes
\begin{equation}
\hat T(\tau)
=
\frac{\tau}{2}\mathbbm 1
+
|T_{01}(\tau)|
\left[
\cos\phi(\tau)\,\sigma_x
+
\sin\phi(\tau)\,\sigma_y
\right].
\label{eq:T_matrix_Bloch}
\end{equation}
The clock readout is binarized at the midpoint \(\tau/2\). The corresponding
binary operator is
\begin{equation}
\hat O(\tau)
=
\operatorname{sgn}
\left[
\hat T(\tau)
-
\frac{\tau}{2}\mathbbm 1
\right].
\label{eq:binarized_operator_def}
\end{equation}
Since
\begin{equation}
\left[
\cos\phi(\tau)\,\sigma_x
+
\sin\phi(\tau)\,\sigma_y
\right]^2
=
\mathbbm 1,
\end{equation}
its eigenvalues are \(\pm1\). Therefore, whenever
$T_{01}(\tau)\neq0$,
the binarized observable is
\begin{equation}
\hat O(\tau)
=
\cos\phi(\tau)\,\sigma_x
+
\sin\phi(\tau)\,\sigma_y .
\label{eq:binarized_curved_observable}
\end{equation}
The magnitude \(|T_{01}(\tau)|\) determines the separation of the two
eigenvalues of the Peres-time operator from the threshold \(\tau/2\), whereas
the phase \(\phi(\tau)\) determines the effective measurement axis on the
equator of the Bloch sphere. The binarized observable depends only on this
phase, provided that the off-diagonal element does not vanish.

\begin{comment}
At the isolated monitoring times
\begin{equation}
\omega_0\tau=2\pi q,
\qquad
q\in\mathbb Z,
\end{equation}
the flat contribution to \(T_{01}\) vanishes. At such points the perturbative
expansion of the phase around the flat result is singular, and the binarized
axis is not defined by the leading flat-space operator. These isolated
monitoring times are excluded from the Bell protocol. The flat calibration
constrains only differences of monitoring times, so the individual monitoring
times can always be chosen away from these zeros.
\end{comment}

\subsection{Weak-curvature expansion of the Bell parameter}
\label{subsec:weak_curvature_CHSH}

We now use the curvature-corrected binarized observables to evaluate the Bell
parameter. The input state is kept fixed throughout the protocol and is chosen to be 
\begin{equation}
|\Phi^+\rangle
=
\frac{|00\rangle+|11\rangle}{\sqrt{2}}.
\label{eq:Phi_plus_fixed}
\end{equation}
For Alice, the two local observables are
\begin{align}
\hat A_a
&=
\cos\phi_a^{\rm curv}\,\sigma_x
+
\sin\phi_a^{\rm curv}\,\sigma_y,\nonumber\\
\hat A_{a'}
&=
\cos\phi_{a'}^{\rm curv}\,\sigma_x
+
\sin\phi_{a'}^{\rm curv}\,\sigma_y.
\end{align}
For Bob, we use the convention
\begin{align}
\hat B_b
&=
\cos\phi_b^{\rm curv}\,\sigma_x
-
\sin\phi_b^{\rm curv}\,\sigma_y,\nonumber\\
\hat B_{b'}
&=
\cos\phi_{b'}^{\rm curv}\,\sigma_x
-
\sin\phi_{b'}^{\rm curv}\,\sigma_y.
\end{align}
With this convention, the correlator for the state \(|\Phi^+\rangle\) is
\begin{equation}
E_{\Phi^+}(\alpha,\beta)
=
\langle\Phi^+|
\hat A_\alpha\otimes\hat B_\beta
|\Phi^+\rangle
=
\cos\left(
\phi_\alpha^{\rm curv}
-
\phi_\beta^{\rm curv}
\right).
\label{eq:Phi_plus_correlator_curved}
\end{equation}
The Bell parameter is therefore
\begin{align}
S_{\Phi^+}^{\rm curv}
&={}
\cos\left(
\phi_a^{\rm curv}
-
\phi_b^{\rm curv}
\right)
+
\cos\left(
\phi_a^{\rm curv}
-
\phi_{b'}^{\rm curv}
\right)
\nonumber\\
&+
\cos\left(
\phi_{a'}^{\rm curv}
-
\phi_b^{\rm curv}
\right)
-
\cos\left(
\phi_{a'}^{\rm curv}
-
\phi_{b'}^{\rm curv}
\right).
\label{eq:CHSH_four_curved_angles}
\end{align}
The four monitoring times are fixed by the flat-spacetime calibration. In flat
spacetime they satisfy
\begin{equation}
\phi_{a'}^{\rm flat}
-
\phi_a^{\rm flat}
=
2\pi n_A,
\qquad
\phi_{b'}^{\rm flat}
-
\phi_b^{\rm flat}
=
2\pi n_B,
\label{eq:flat_phase_calibration_CHSH}
\end{equation}
or, equivalently,
\begin{equation}
\tau_{a'}
-
\tau_a
=
\frac{4\pi n_A}{\omega_0},
\qquad
\tau_{b'}
-
\tau_b
=
\frac{4\pi n_B}{\omega_0}.
\label{eq:flat_time_calibration_CHSH}
\end{equation}
The unprimed flat settings are chosen to coincide,
\begin{equation}
\phi_a^{\rm flat}
-
\phi_b^{\rm flat}
=
0
\qquad
(\mathrm{mod}\ 2\pi).
\label{eq:flat_reference_axes}
\end{equation}
It follows that the flat spacetime bound is saturated,
\begin{equation}
S_{\Phi^+}^{\rm flat}=2.
\end{equation}
We write the four curved phases as
\begin{align}
\phi_a^{\rm curv}
&=
\phi_a^{\rm flat}
+
\delta\phi_A(\tau_a),\nonumber\\
\phi_{a'}^{\rm curv}
&=
\phi_{a'}^{\rm flat}
+
\delta\phi_A(\tau_{a'}),\nonumber\\
\phi_b^{\rm curv}
&=
\phi_b^{\rm flat}
+
\delta\phi_B(\tau_b),\nonumber\\
\phi_{b'}^{\rm curv}
&=
\phi_{b'}^{\rm flat}
+
\delta\phi_B(\tau_{b'}).
\end{align}
The curvature-induced splittings of the local settings are
\begin{align}
\epsilon_A
&=
\delta\phi_A(\tau_{a'})
-
\delta\phi_A(\tau_a),\nonumber\\
\epsilon_B
&=
\delta\phi_B(\tau_{b'})
-
\delta\phi_B(\tau_b).
\label{eq:epsilon_curvature_phase}
\end{align}
The curvature may also produce a relative shift between the two unprimed
reference axes. We denote this by
\begin{equation}
\xi
=
\delta\phi_A(\tau_a)
-
\delta\phi_B(\tau_b)=\phi_a^{\rm curv}-\phi_b^{\rm curv}.
\label{eq:common_axis_mismatch}
\end{equation}
Using Eqs.~\eqref{eq:flat_phase_calibration_CHSH} and
\eqref{eq:flat_reference_axes}, the four phase differences entering
Eq.~\eqref{eq:CHSH_four_curved_angles} become, modulo \(2\pi\),
\begin{align}
\phi_a^{\rm curv}
-
\phi_b^{\rm curv}
&=
\xi,\nonumber\\
\phi_a^{\rm curv}
-
\phi_{b'}^{\rm curv}
&=
\xi-\epsilon_B,\nonumber\\
\phi_{a'}^{\rm curv}
-
\phi_b^{\rm curv}
&=
\xi+\epsilon_A,\nonumber\\
\phi_{a'}^{\rm curv}
-
\phi_{b'}^{\rm curv}
&=
\xi+\epsilon_A-\epsilon_B.
\end{align}
Consequently,
\begin{align}
S_{\Phi^+}^{\rm curv}
={}&
\cos\xi
+
\cos(\xi-\epsilon_B)
+
\cos(\xi+\epsilon_A)
\nonumber\\
&-
\cos(\xi+\epsilon_A-\epsilon_B).
\label{eq:CHSH_xi_epsilon_exact}
\end{align}
In the weak-curvature regime,
\begin{equation}
|\xi|\ll1,
\qquad
|\epsilon_A|\ll1,
\qquad
|\epsilon_B|\ll1,
\end{equation}
this gives
\begin{equation}
S_{\Phi^+}^{\rm curv}
=
2
-
\xi^2
-
\epsilon_A\epsilon_B
+
\mathcal O(R^3).
\label{eq:CHSH_weak_general}
\end{equation}
Here \(\xi\), \(\epsilon_A\), and \(\epsilon_B\) are all first order in the
curvature. The leading change in the CHSH value is therefore quadratic in the
weak-curvature scale.

The relative unprimed-axis shift can be eliminated by a symmetric choice of
the reference monitoring times. For example, if the two particles
experience the same curvature profile (and they do for constant curvature) and $\tau_a=\tau_b,
\label{eq:symmetric_unprimed_times}$, 
then $\delta\phi_A(\tau_a)
=
\delta\phi_B(\tau_b)$,
and hence 
$\xi=0$.
Equation~\eqref{eq:CHSH_weak_general} then reduces to
\begin{equation}
S_{\Phi^+}^{\rm curv}
=
2
-
\epsilon_A\epsilon_B
+
\mathcal O(R^3).
\label{eq:CHSH_weak_symmetric}
\end{equation}
Thus, the Bell state \(|\Phi^+\rangle\) violates the CHSH inequality whenever the following condition holds,
\begin{equation}
\epsilon_A\epsilon_B<0.
\label{eq:fixed_state_violation_condition}
\end{equation}
\vspace{1pt}

For constant curvature, the curvature correction to the effective Peres-time
angle is given by Eq.~\eqref{eq:integrated_effective_angle_R}. Using the
flat-space calibration
\begin{equation}
\tau_{a'}
-
\tau_a
=
\frac{4\pi n_A}{\omega_0},
\qquad
\tau_{b'}
-
\tau_b
=
\frac{4\pi n_B}{\omega_0},
\end{equation}
the oscillatory part of Eq.~\eqref{eq:integrated_effective_angle_R} cancels
from the local differences. Hence
\begin{align}
\epsilon_A
&=
-\frac{\pi n_A R_A}{2\omega_0^2}
+
\mathcal O\left(
\frac{R_A^2}{\omega_0^4}
\right),\label{eq:epsilon_A_scalar_curvature}\\
\epsilon_B
&=
-\frac{\pi n_B R_B}{2\omega_0^2}
+
\mathcal O\left(
\frac{R_B^2}{\omega_0^4}
\right).
\label{eq:epsilon_B_scalar_curvature}
\end{align}
Substituting~\eqref{eq:epsilon_A_scalar_curvature} and
\eqref{eq:epsilon_B_scalar_curvature} into~\eqref{eq:CHSH_weak_symmetric}, one obtains
the curved Bell parameter 
\begin{equation}
S_{\Phi^+}^{\rm curv}
=
2
-
\frac{\pi^2 }{4\omega_0^4}
n_A n_B R_A R_B
+
\mathcal O(R^3).
\label{eq:CHSH_scalar_curvature}
\end{equation}
The violation condition is therefore
\begin{equation}
n_A n_B R_A R_B<0.
\label{eq:violation_scalar_condition}
\end{equation}
The sign of \(n_A\) or \(n_B\) is fixed by the ordering of the two monitoring
times at the corresponding laboratory. For instance, \(n_A<0\) means
$\tau_{a'}<\tau_a$,
while both monitoring times themselves remain positive. 

For (A)dS\(_2\),
\begin{equation}
R_A=R_B=\pm 2/\ell^2.
\end{equation}
Equation~\eqref{eq:CHSH_scalar_curvature} gives
\begin{equation}
S_{\Phi^+}^{\rm AdS}
=
2
-
\frac{\pi^2}{\ell^4\omega_0^4}n_A n_B
+
\mathcal O\left(
\frac{1}{\ell^6\omega_0^6}
\right).
\label{eq:CHSH_AdS_fixed_state}
\end{equation}
If the two local calibrations are chosen with opposite orientation,
$n_A n_B<0$,
this becomes
\begin{equation}
S_{\Phi^+}^{\rm (A)dS}
=
2
+
\frac{\pi^2 }{\ell^4\omega_0^4}|n_A n_B|
+
\mathcal O\left(
\frac{1}{\ell^6\omega_0^6}
\right)
>
2.
\label{eq:CHSH_AdS_violation}
\end{equation}
For dS\(_2\), 
the same expression applies provided that the trap remains stable, $
\omega_0^2>\frac{1}{\ell^2}$. 
Thus, AdS\(_2\) and dS\(_2\) differ in the sign of the frequency shift, but for
equal curvatures on the two sides, the sign of Bell activation
is controlled by the relative orientation of the two local monitoring-time
calibrations. In both cases, for \(R_A R_B>0\), the fixed state
\(|\Phi^+\rangle\) violates the CHSH inequality when $
n_A n_B<0$.

\end{document}